\documentclass[12 pt]{article}
\RequirePackage[numbers]{natbib}
\RequirePackage[colorlinks,citecolor=blue,urlcolor=blue]{hyperref}

\usepackage{amssymb}
\usepackage[all,knot,arc,import,poly]{xy}
\usepackage{graphicx}
\usepackage{amsmath}
\usepackage{color}
\usepackage{enumerate}
\usepackage{hyperref}
\usepackage{tikz}
\usetikzlibrary{positioning,shapes,arrows, snakes,fit}
\usetikzlibrary{matrix}
\usepackage{graphicx}
\usepackage{subfig}

\newcommand{\beginsupplement}{%
        \setcounter{table}{0}
        \renewcommand{\thetable}{S\arabic{table}}%
        \setcounter{figure}{0}
        \renewcommand{\thefigure}{S\arabic{figure}}%
     }

\begin{document}
\begin{center}{
\textbf{\Large Sequential double cross-validation for assessment of added predictive ability in high-dimensional omic applications}\\
}
\end{center}

Mar Rodr\'{i}guez-Girondo$^{1}$, Perttu Salo$^{2}$, Tomasz Burzykowski$^{3}$, Markus Perola$^{2}$, Jeanine Houwing-Duistermaat$^{1,4}$ and Bart Mertens$^{1}$
\bigskip

\noindent
$^{1}$Department of Medical Statistics and Bioinformatics, Leiden University Medical Center, Leiden, The Netherlands.

\noindent
$^{2}$National Institute For Health and Welfare, Helsinki, Finland.

\noindent
$^{3}$Interuniversity Institute for Biostatistics and statistical Bioinformatics (I-BioStat), Hasselt University, Hasselt, Belgium.

\noindent
$^{4}$Department of Statistics, Leeds University, Leeds, United Kingdom.

\begin{abstract}
Enriching existing predictive models with new biomolecular markers is an important task in the new multi-omic era. Clinical studies increasingly include new sets of omic measurements which may prove their added value in terms of predictive performance. We introduce a two-step approach for the assessment of the added predictive ability of omic predictors, based on sequential double cross-validation and regularized regression models. We propose several performance indices to summarize the two-stage prediction procedure and a permutation test to formally assess the added predictive value of a second omic set of predictors over a primary omic source. The performance of the test is investigated through simulations. We illustrate the new method through the systematic assessment and comparison of the performance of transcriptomics and metabolomics sources in the prediction of body mass index (BMI) using longitudinal data from the Dietary, Lifestyle, and Genetic determinants of Obesity and Metabolic syndrome (DILGOM) study, a population-based cohort from Finland.
\end{abstract}

\section{Introduction\label{section intro}}
During the past decade, much attention has been devoted to accommodate single high-dimensional sources of molecular data (omics) in the calibration of prediction models for health traits. For example, microarray-based transcriptome profiling and mass spectometry proteomics have been established as promising omic predictors in oncology \cite{Dudoit, Rosenwald, Schwamborn} and, to lesser extent, in  metabolic health \cite{Apalasamy, Jenkinson}. Nowadays, due to technical advances in the field and evolving biological knowledge, novel omic measures, such as NMR proteomics and metabolomics \cite{Inouye1, Stroeve} or  nano-LCMS and UPLC glycomics \cite{Zoldos, Theodoratou} are emerging as potentially powerful new biomolecular marker sets. As a result, it is increasingly common for studies to collect a range of omic measurements in the same set of individuals, using different measurement platforms and covering different aspects of human biology. This causes new statistical challenges, among which the evaluation of the ability of novel biomolecular markers to improve predictions based on previously established predictive omic sources, often referred as their added or incremental predictive ability \cite{Pencina1, Pencina2, Hilden}.  



An illustrative example of these new methodological challenges is given by our motivating problem. We have access to data from 248 individuals sampled from the Helsinki area, Finland, within the Dietary, Lifestyle, and Genetic determinants of Obesity and Metabolic syndrome (DILGOM) study \cite{Inouye1}. One-hundred-thirty-seven highly correlated NMR serum metabolites and 7380 beads from array-based transcriptional profiling from blood leukocytes were measured at baseline in 2007, together with a large number of clinical and demographic factors which were also measured in 2014, after seven years of follow-up. Our primary goal is the prediction of future body mass index (BMI), using biomolecular markers measured at baseline. More specifically, we would like to compare the predictive ability of the available metabolomics and transcriptomics, and to determine if both should be retained in order to improve single-omic source predictions of BMI at seven years of follow-up.

In our setting, it is necessary to both calibrate the predictive model based on each source of omic predictors and assess the incremental predictive ability of a secondary one relative to the first set, using the same set of observations. Hence, in order to avoid overoptimism and provide realistic estimates of performance, it is necessary to control for the re-use of the data, which has already been employed for model fitting within the same observations \cite{Stone, Jonathan, Varma}. This is a very important issue in omic research, where external validation data are hard to obtain. It is well known that biased estimation of model performance due to re-use of the data increases with large number of predictors \cite{Hastie1} and omic sets are typically high-dimensional ($n<p$, $n$ sample size and $p$ the number of predictors). Extra difficulties in our setting are the different dimensions (number of features), scales and correlation structure of each omic source, and possible correlation between omic sources induced by partially common underlying biological information. 

Evaluating the added predictive ability of new biomarkers regarding classical, low dimensional, settings has been a topic of intense debate in the biostatistical literature in the last years (see, for example, \cite{Pencina2, Hilden, Kerr, Pepe} and references therein). Getting meaningful summary measures and valid statistical procedures for testing the added predictive value are difficult tasks, even when considering the addition of a single additional biomarker in the classical regression context. In particular, widely used testing procedures for improvement in discrimination based on area under the ROC curve (AUC) differences \cite{deLong} and net reclassification index (NRI) \cite{Pencina1} have shown unacceptable false positive rates in recent simulation studies \cite{Kerr,Pepe}. Overfitting is a big problem when comparing estimated predictions coming from nested regression models fitted in the same dataset. Moreover, the distributional assumptions of the proposed tests seem inappropriate, translating into poor performance of the aforementioned tests even when using independent validation sets \cite{Pepe}.

To date, little attention has been given to the evaluation of the added predictive ability in high-dimensional settings, where the aforementioned problems are larger and new ones appear, such as the simultaneous inclusion in an unique prediction model of predictors sets of very different nature. Tibshirani and Efron \cite{preval1} have shown that overfitting may dramatically inflate the estimated added predictive ability of omic sources with respect to a low-dimensional set of clinical parameters. To solve this issue, they proposed to first create a univariate `pre-validated' omic predictor based on cross-validation techniques \cite{Stone, Jonathan,Breiman, Mertens1, Mertens2} and incorporate it as a new covariate to the regression with low-dimensional clinical parameters. In a subsequent publication, Hoefling and Tibshirani \cite{preval2} have shown that standard tests in regression models are biased for pre-validated predictors. As a solution, the authors suggest a permutation test which seems to perform well under independence of clinical and omic sets. Boulesteix and Hothorn \cite{boosting} have proposed an alternative method for the same setting of enriching clinical models with a high-dimensional set of predictors. In contrast to \cite{preval1,preval2}, they first obtain a clinical prediction based on traditional regression techniques. In a second step, the clinical predictor is incorporated as an offset term in a boosting algorithm based on the omic source of predictors. Previous calibration of the clinical prediction is not addressed in the second step and the same permutation strategy than Hoefling and Tibshirani \cite{preval2} is used to derive p-values.   


In this paper, we propose a two-step procedure for the assessment of additive predictive ability regarding two high-dimensional and correlated sources of omic predictors. To the best of our knowledge, no previous work has addressed this problem before. Our approach combines double cross-validation sequential prediction based on regularized regression models and a permutation test to formally assess the added predictive value of a second omic set of predictors over a primary omic source.

\section{Methods}
Let the observed data be given by $(\mathbf{y},\mathbf{X}_{1},\mathbf{X}_{2})$, where $\mathbf{y}=(y_{1},\ldots,y_{n})^{\intercal}$ is the continuous outcome measured in $n$ independent individuals and $\mathbf{X}_{1}$ and $\mathbf{X}_{2}$ are two matrices of dimension $n\times p$ and  $n \times q$, respectively, representing two omic predictor sources with $p$ and $q$ features. We assume that we are in a high-dimensional setting ($p,q>n$) and that the main goal is to evaluate the incremental or added value of $\mathbf{X}_2$ beyond $\mathbf{X}_1$ in order to predict $\mathbf{y}$ in new observations. Our approach is based on comparing the performance of a primary model based only on $\mathbf{X}_{1}$ with an extended model based on $\mathbf{X}_{2}$ and adjusted by the primary fit based on $\mathbf{X}_{1}$.

\subsection{Sequential estimation with two sources of predictors}
We propose a two-step  procedure based  on the replacement of the original (high-dimensional) sources of predictors by their corresponding estimated values of $\mathbf{y}$ based on a single-source-specific prediction model.

In the first step, we build a prediction model for $\mathbf{y}$ based on $\mathbf{X}_{1}$ and a given model specification $f$. Based on the fitted model, the fitted values $\mathbf{p}_{1}=\hat{f}(\mathbf{X}_{1})=(p_{11},\ldots,p_{1n})^{\intercal}$ are estimated. Then, for each individual $i$, we take the residual $res_{i}=y_{i}-p_{1i}$. We consider $\mathbf{res}=(res_{1},\ldots,res_{n})^{\intercal}$ as new response and construct a second prediction model based on $\mathbf{X}_2$ as predictor source:
\begin{equation}
\bf{p}_{2}=E(\mathbf{res}|\mathbf{X}_{2})=f(\mathbf{X}_{2})
\end{equation}
This is equivalent to including $\mathbf{p}_{1}$ as an offset term (fixed) in the model based on $\mathbf{X_{2}}$ for the prediction of the initial outcome $\mathbf{y}$:
\begin{equation}
\bf{p}_{2}=E(\mathbf{y}|\mathbf{X}_{2},\bf{p}_{1})=f(\mathbf{X}_{2})+\bf{p_{1}}
\end{equation}

Several statistical methods are available to derive prediction models of continuous outcomes in high-dimensional settings. In this work, we focus on regularized linear regression models \cite{Hastie1},
where $f(\mathbf{X})=\mathbf{X}\boldsymbol \beta$ and 
the estimation of $\boldsymbol{\beta}$ is conducted by solving $min_{\boldsymbol \beta}(\mathbf{X}\boldsymbol \beta-\mathbf{Y})^{\intercal}(\mathbf{X}\boldsymbol \beta-\mathbf{Y})+\lambda pen(\boldsymbol \beta)$, where $pen(\boldsymbol \beta)=\frac{1-\alpha}{2}||\boldsymbol \beta||_{2}^{2}+\alpha||\boldsymbol \beta||_{1}$. The penalty parameter $\lambda$ regularizes the $\boldsymbol \beta$ coefficients, by shrinking large coefficients in order to control the bias-variance trade-off. The pre-fixed parameter $\alpha$ determines the type of imposed  penalization. We used two widely used penalization types: $\alpha=0$ (ridge, i.e., \ $\ell_{2}$ type penalty \cite{Hoerl}) and $\alpha=1$ (lasso, i.e., \ $\ell_{1}$ penalty \cite{Tibshirani1}). Note that other  model building strategies for prediction of continuous outcomes could have been used in this framework, such as the elastic net penalization \cite{Zou} by setting $\alpha=0.5$, or  boosting methods \cite{Tutz, Buhlmann, Kneib}, among others.

\subsection{Double cross-validation prediction}
The use of a previously estimated quantity ($\bf{p}_{1}$) in the calibration of a prediction model based on $\mathbf{X}_2$ (expressions (1) and (2)) requires, in absence of external validation data, the use of resampling techniques to avoid bias in the assessment of the role of $\bf{p}_{1}$ and $\bf{p}_{2}$. We use double cross-validation algorithms \cite{preval1,Mertens1, Mertens2,preval2}, consisting of two nested loops. In the inner loop a cross-validated grid-selection is used to determine the optimal prediction rule, i.e., for model selection, while the outer loop  is used to estimate the prediction performance by application of models developed in the inner loop part of the data (training sets) to the remaining unused data (validation sets). In this manner, double cross-validation is capable of avoiding the bias in estimates of predictive ability which would result from use of a single-cross-validatory approach only. In our setting, the outer loop of a `double' cross-validatory calculation allows obtaining `predictive'-deletion residuals which fully account for the inherent uncertainty of model fitting on the primary source ($\mathbf{X}_{1}$),  before assessing the added predictive ability of $\mathbf{X}_{2}$, given by $\bf{p}_{2}$. The basic structure of the double cross-validation procedure to estimate unbiased versions of $\bf{p}_{1}$ and $\bf{p}_{2}$ is as follows:

\begin{description}
\item[Step 1]
Obtain  double cross-validation predictions of $\mathbf{y}$, $\mathbf{p}_{1}=(p_{11},\ldots,p_{1n})$, based on $\mathbf{X_{1}}$:
\begin{description}
\item[-] Randomly split sample $S$ in $J$ mutually exclusive and exhaustive sub-samples of approximately equal size $S^{(1)},\ldots, S^{(J)}$
\item[-] For each $j=1,\ldots,J$, merge $J-1$ subsamples into $S^{(-j)}=S-S^{(j)}$
\begin{description}
\item[-] Randomly split sample  $S^{(-j)}$ in $K$ sub-samples $(S^{(-j)})^{(1)},\ldots,(S^{(-j)})^{(K)}$
\item[-] For each $k=1,\ldots,K$, merge $K-1$ subsubsamples into $(S^{(-j)})^{(-k)}=S^{(-j)}-(S^{(-j)})^{(k)}$
\begin{description} 
\item[*] Fit regression model $\mathbf{y}=\widehat{f}_{\lambda_{l}}^{(-k)}(\mathbf{X}_{1})+\boldsymbol\epsilon$ for a grid of values of shrinkage parameters $\lambda_l$, $l=1,\ldots,L$ to $(S^{(-j)})^{(-k)}$
\item[*] Evaluate $\widehat{f}_{\lambda_{l}}^{(-k)}$,$l=1,\ldots,L$ in the $kth$ held-out sub-sample $(S^{(-j)})^{(k)}$ by calculating $\widehat{e}_{\lambda_{l}}^{k}=\sum_{(\mathbf{y},\mathbf{X}_{1})\in (S^{(-j)})^{(k) } }(\mathbf{y}-\widehat{f}^{(-k)}_{\lambda_{l}}(\mathbf{X}_{1}))^2$
\end{description}
\item[-] Compute overall cross-validation error: $\widehat{e}_{\lambda_{l}}^{(-j)}=\frac{1}{K}\sum_{k=1}^{K}\widehat{e}_{\lambda_{l}^{k}}$, $l=1,\ldots,L$
\item[-] Choose $\lambda_{opt}^{(-j)}=min_{l=1,\ldots,L}(\widehat{e}_{\lambda_{l}}^{(-j)})$ and calculate predictions of $\mathbf{y}$ in the $jth$ held-out sub-sample $S^{(j)}$, $\mathbf{p}_{1}^{(j)}=\widehat{f}_{\lambda_{opt}^{(-j)}}(\mathbf{X}_{1})$, $(\mathbf{y},\mathbf{X}_{1})\in S^{(j)}$
\end{description}
\item[-] The vector of predictions of $\mathbf{y}$, $\mathbf{p}_{1}=(p_{11},\ldots,p_{1n})$ is obtained by concatenating the $J$ $\mathbf{p}_{1}^{(j)}$, $j=1,\ldots,J$ vectors, i.e., $\mathbf{p}_{1}=(\mathbf{p}_{1}^{(1)},\ldots,\mathbf{p}_{1}^{(J)})$
\end{description}
\item[Step 2] Repeat the process detailed in Step 1 considering the double cross-validated residuals $\mathbf{res}=(y_{1}-p_{11},\ldots,y_{n}-p_{1n})$, as outcome and $\mathbf{X_{2}}$ as set of predictors and obtain the double cross-validation predictions $\mathbf{p}_{2}=(p_{21},\ldots,p_{2n})^{\intercal}$. Note that this is equivalent to obtaining the double cross-validation predictions of $\mathbf{y}$ based on $\mathbf{X_{2}}$ considering $\mathbf{p}_{1}$ as offset variable in the $J$ fits of model (1).
\end{description}

\subsection{Summary measures of predictive accuracy based on double cross-validation}
In order to evaluate the performance of the sequential procedure introduced in Subsection 2.1., we propose three measures of predictive accuracy, denoted by $Q^{2}_{\mathbf{X}_1}$, $Q^{2}_{\mathbf{X}_2|\mathbf{X}_1}$, and $Q^{2}_{\mathbf{X}_{1},\mathbf{X}_{2}}$, based on sum of squares of the double cross-validated predictions $\bf{p}_{1}$ and $\bf{p}_{2}$, obtained following the procedure described in Subsection 2.2.. These summary measures can be regarded as high-dimensional equivalents of calibration measurements for continuous outcomes in low-dimensional settings \cite{Schemper, Westerhuis}, and an extension of previously discussed proposals in the cross-validation literature \cite{Jonathan}.

Denote by $PRESS(\mathbf{y},\mathbf{p})=\sum_{i=1}^{n}(y_{i}-p_{i})^{2}$ the prediction sum of squares based on a vector of predictions $\mathbf{p}$, obtained according to some arbitrary model $f$, $\mathbf{p}=(p_{1},\ldots,p_{n})^{\intercal}=E(\mathbf{y}|\mathbf{X})$ and by $CVSS(\mathbf{p}_{1},\mathbf{p}_{2})=\sum_{i=1}^{n}(p_{1i}-p_{2i})^{2}$ the sum of squared differences between two cross-validated vectors of predictions, e.g. $\mathbf{p}_{1}=E_{f_{1}}(\mathbf{y}|\mathbf{X}_{1})$, $\mathbf{p}_{2}=E_{f_{2}}(\mathbf{y}|\mathbf{X}_{2})$. Let $\mathbf{p}_{0}$ be the simplest cross-validated predictor of $\mathbf{y}$, based on the sample mean of $\mathbf{y}$ only. To summarize the first step of the sequential procedure, we use double cross-validation to estimate the predictive ability of $\mathbf{X}_{1}$ by

\begin{equation}
Q_{\mathbf{X}_{1}}^{2}=\frac{CVSS(\mathbf{p_{1}},\mathbf{p_{0}})}{PRESS(\mathbf{y},\mathbf{p_{0}})}=\frac{\sum_{j=1}^{J}\sum_{j\in S^{(j)}}\left(p_{1j}-\overline{y}^{(-j)}\right)^2}{\sum_{j=1}^{J}\sum_{j\in S^{(j)}}\left(y_{j}-\bar{y}^{(-j)}\right)^2}.
\end{equation}

Intuitively, $Q_{\mathbf{X}_{1}}^2$ represents the proportion of the variation of the response $\mathbf{y}$ that is expected to be explained by $f(\mathbf{X}_{1})$ in new individuals , re-scaled by the total amount of prediction variation in the response $\mathbf{y}$. In the worst case scenario, when $\mathbf{p}_{1}=\mathbf{p}_{0}$ ($\mathbf{X}_{1}$ as predictive as a null model based on the mean of $\mathbf{y}$) $Q_{\mathbf{X}_{1}}^2=0$ and $Q_{\mathbf{X}_{1}}^2=1$ if $\mathbf{p}_{1}=\mathbf{y}$. Since the computation of $p_{1j}$, $j\in S^{(j)}$ for each of the $j=1,\ldots,J$ random splits of the sample $S$ is based on the observations not belonging to $S^{(j)}$, we proceed in an analogous way to compute the average predicted variation of $\mathbf{y}$. Hence, in order to get an appropriate re-scaling factor, for each subset $S^{(j)}$, we compute $\bar{y}^{(-j)}$, the mean value of the outcome variable $\mathbf{y}$ calculated without the observations belonging to $S^{(j)}$. 

Assume that $Q_{\mathbf{X_{1}}}^{2}>0$, the contribution of the second omic source, $\mathbf{X}_{2}$, in the prediction of $\mathbf{y}$ can be summarized by 

\begin{equation}
Q_{\mathbf{X}_{2}|\mathbf{X}_{1}}^{2}=\frac{CVSS(\mathbf{p_{2}},\overline{\mathbf{y}-\mathbf{p_{1}})}}{PRESS(\mathbf{y}-\mathbf{p_{1}},\overline{\mathbf{y}-\mathbf{p_{1}}})}=\frac{\sum_{j=1}^{J}\sum_{j\in S^{(j)}}\left(p_{2j}-\overline{(y_{j}-p_{1j})}^{(-j)}\right)^2}{\sum_{j=1}^{J}\sum_{j\in S^{(j)}}\left(y_{j}-p_{1j}-\overline{(y_{j}-p_{1j})}^{(-j)}\right)^2}.
\end{equation}

$Q_{\mathbf{X}_{2}|\mathbf{X}_{1}}^{2}$ accounts for the predictive capacity of $f(\mathbf{X}_{2})$, after removing the part of variation in $\mathbf{y}$ that can be attributed to the first source of predictors $\mathbf{X}_{1}$. Its computation relies on the squared difference between $\mathbf{p}_2$ (the double cross-validated predictions resulting from the second step of the proposed procedure in Subsection 2.2.) and the corresponding residual from the step 1 ($\mathbf{res}=\mathbf{y}-\mathbf{p}_{1}$) based on $\mathbf{X}_{1}$, re-scaled by the remaining predicted variation on $\mathbf{y}$ after the first step of the procedure. As a result, $Q_{\mathbf{X}_{2}|\mathbf{X}_{1}}^{2}$ can be regarded as the expected ability of $\mathbf{X}_{2}$ to predict the part of $\mathbf{y}$, after adjusting for the predictive capacity of $f(\mathbf{X}_{1})$ and accounting for all model fitting in the first stage of the assessment. Following the same arguments used in deriving expression (3), for each subset $S^{(j)}$, the computation of the average variation of $\mathbf{res}$  is based on $\overline{(y-p_{1})}^{(-j)}$, i.e., excluding the observations belonging to $S^{(j)}$. Note that in Step 1 of the sequential procedure $J$ models are fitted, each based on $S^{(-j)}$, providing residuals with expected zero mean (given specification (1)), i.e., $\overline{(y-p_{1})}^{(-j)}\approx0$, $j=1,\ldots,J$. Hence, ${\sum_{j=1}^{J}\sum_{j\in S^{(j)}}\left(y_{j}-p_{1j}-\overline{(y-p_{1})}^{(-j)}\right)^2}\approx \sum_{i=1}^{n}\left(y_{i}-p_{1i}\right)^2$ and thus
\begin{equation}
Q_{\mathbf{X}_{2}|\mathbf{X}_{1}}^{2}\approx\frac{\sum_{i=1}^{n}p_{2i}^2}{\sum_{i=1}^{n}\left(y_{i}-p_{1i}\right)^2}.
\end{equation}
Finally, we derive a third summarizing measurement of the overall sequential process, $Q_{\mathbf{X}_{1},\mathbf{X}_{2}}^{2}$, defined as: 
\begin{equation}
Q_{\mathbf{X}_{1},\mathbf{X}_{2}}^2=\frac{CVSS(\mathbf{p_{1}}+\mathbf{p_{2}},\mathbf{p_{0}})}{PRESS(\mathbf{y},\mathbf{p_{0}})}=\frac{\sum_{j=1}^{J}\sum_{j\in S^{(j)}}\left(p_{1j}+p_{2j}-\bar{y}^{(-j)}\right)^2}{\sum_{j=1}^{J}\sum_{j\in S^{(j)}}\left(y_{j}-\bar{y}^{(-j)}\right)^2}.
\end{equation}
$Q_{\mathbf{X}_{1},\mathbf{X}_{2}}^2$ represents the total predictive capacity of the overall sequential procedure based on $\mathbf{X}_{1}$ and $\mathbf{X}_{2}$, i.e., the combined predictive ability of $\mathbf{X}_{1}$ and $\mathbf{X}_{2}$ given by $\bf{p}_{1}+\bf{p}_{2}$. Note that $Q_{\mathbf{X}_{1},\mathbf{X}_{2}}^2$ is based on the same squared difference between $\mathbf{p}_2$ and $\mathbf{res}=\mathbf{y}-\mathbf{p}_{1}$ as $Q_{\mathbf{X}_{2}|\mathbf{X}_{1}}^{2}$, but the re-scaling factor refers to the total predictive variation of the original response $\mathbf{y}$.

The three introduced measures jointly summarize the  performance of the two omic sources under study and their interplay in order to predict the outcome $\mathbf{y}$. In all the cases higher values are indicative of higher predictive ability. The three measurements vary between 0 (null predictive ability) and the maximal value of 1. The interpretation of $Q^{2}_{\mathbf{X}_1}$ is straightforward, as it simply captures the predictive capacity of the firstly evaluated omic source. Note that the difference between $Q^{2}_{\mathbf{X}_2|\mathbf{X}_1}$ and $Q^{2}_{\mathbf{X}_{1},\mathbf{X}_{2}}$ relies on the denominator. In general, if $\mathbf{X}_{1}$ is informative, the denominator in expression (4) will be smaller than in expression (6). Thus, the residual variation after Step 1 will be smaller than the total initial variation. 

The three summary measures are related by the following expression:
\begin{equation}
(1-Q_{\mathbf{X}_1}^2)(1-Q_{\mathbf{X}_2|\mathbf{X}_1}^2)\approx(1-Q_{\mathbf{X}_{1},\mathbf{X}_{2}}^2).
\end{equation}

Consequently, we can rewrite $Q_{\mathbf{X}_2|\mathbf{X}_1}^2$ as follows:
\begin{equation}
Q_{\mathbf{X}_2|\mathbf{X}_1}^2\approx\frac{Q_{\mathbf{X}_{1},\mathbf{X}_{2}}^{2}-Q_{\mathbf{X}_{1}}^{2}}{(1-Q_{\mathbf{X}_{1}}^2)}.
\end{equation}

Note that in cases in which $Q_{\mathbf{X}_{2}|\mathbf{X}_{1}}^{2}=0$, we get that $Q^{2}_{\mathbf{X}_{1},\mathbf{X}_{2}}-Q^{2}_{\mathbf{X}_{1}}=0$, and viceversa. However,  $Q_{\mathbf{X}_{2}|\mathbf{X}_{1}}^{2}$ and $Q^{2}_{\mathbf{X}_{1},\mathbf{X}_{2}}-Q^{2}_{\mathbf{X}_{1}}$ differ when not zero. Specifically, from expression (8), we obtain that $Q_{\mathbf{X}_{2}|\mathbf{X}_{1}}^{2}\geq Q^{2}_{\mathbf{X}_{1},\mathbf{X}_{2}}-Q^{2}_{\mathbf{X}_{1 }}$. In short, $Q_{\mathbf{X}_{2}|\mathbf{X}_{1}}^{2}$ may be regarded as the conditional contribution of $\mathbf{X}_{2}$ for the prediction of $\mathbf{y}$ with respect to what may be predicted using $\mathbf{X}_{1}$ alone. $Q_{\mathbf{X}_{1},\mathbf{X}_{2}}^{2}-Q_{\mathbf{X_{1}}}^{2}$ measures the absolute gain  in predictive ability from adding $\mathbf{X_{2}}$  to $\mathbf{X}_{1}$. Note that a given source $\mathbf{X}_{2}$ may present a large conditional $Q_{\mathbf{X}_{2}|\mathbf{X}_{1}}^{2}$  but a small absolute $Q_{\mathbf{X}_{1},\mathbf{X}_{2}}^{2}-Q_{\mathbf{X_{1}}}^{2}$ (if, for example, $\mathbf{X}_{1}$ presents high predictive ability itself). Moreover, due to the relation between $\mathbf{p}_{1}$ and the resulting vector of predictions after combining $\mathbf{X}_{1}$ and $\mathbf{X}_{2}$, $\bf{p}_{1}+\bf{p}_{2}$, expression (8) implies that $Q_{\mathbf{X}_{1},\mathbf{X}_{2}}^{2}\geq Q_{\mathbf{X}_{1}}^{2}$. This desirable property may not be fulfilled using alternative combination strategies. 


In practice, our sequential procedure relies on the realistic assumption of positive predictive ability of the first source of predictors, $\mathbf{X}_{1}$ (one would only be interested in assessing additional or incremental information on top of an informative source itself). Accordingly, we advise to conduct our sequential procedure using $\mathbf{X}_{1}$ as primary source only if $Q_{\mathbf{X}_{1}}^{2}>0$, which is, furthermore, required to derive expression (8).

\subsection{Permutation test for assessment of added predictive ability}
The summary measures may be used to introduce formal tests for assessing the added or augmented predictive value of   $\mathbf{X}_{2}$ over $\mathbf{X}_{1}$ to predict $\mathbf{y}$. We propose a permutation procedure to test the null hypothesis $H_{0}:Q_{\mathbf{X}_{2}|\mathbf{X}_{1}}^2=0$ against the alternative hypothesis $H_{1}:Q_{\mathbf{X}_{2}|\mathbf{X}_{1}}^2>0$. The test is based on permuting the residuals obtained after applying the first step of our two-stage procedure with the data at hand. Our goal is to remove the potential association between $\mathbf{X}_2$ and $\mathbf{y}$ while preserving the original association between $\mathbf{y}$ and $\mathbf{X}_1$.

Explicitly, we propose the following algorithm:

\begin{description}
\item[Step 1] 
Calculate the residuals $\mathbf{res}=\mathbf{y}-\mathbf{p_{1}}$ based on the predictions $\mathbf{p_{1}}$ of $\mathbf{y}$ based on $\mathbf{X}_{1}$, obtained in the first step of the procedure presented in Section 2.1.
\item[Step 2]
Permute the values of $\mathbf{res}$, obtaining $\mathbf{res}^{\pi}$ and generate values of the response $\mathbf{y}$ under the null hypothesis: $\mathbf{y}^{*}=\mathbf{p_{1}}+\mathbf{res}^{\pi}$.
\item[Step 3]
Repeat the two-stage procedure from Section 2.1. for predicting $\mathbf{y}^{*}$ and obtain the corresponding $Q_{\mathbf{X}_{2}|\mathbf{X}_{1}}^{2*}$.
\end{description}

The procedure is repeated $M$ times and the resulting permutation p-values are obtained as follows:
$$
\text{p-value}=\frac{1}{M}\sum_{i=1}^{M}I(Q_{\mathbf{X}_{2}|\mathbf{X}_{1}}^{2*i}>Q_{\mathbf{X}_{2}|\mathbf{X}_{1}}^2),
$$
where $M$ is the number of permutations, and $Q_{\mathbf{X}_{2}|\mathbf{X}_{1}}^{2}$ is the actual observed value with the data at hand. Note that in {\bf Step 2}, we generate a `null' version of the original response $\mathbf{y}$ and then we repeat the overall two-stage procedure, which implies that the `null' residuals used in {\bf Step 3} are not fixed and are, in general, different from $\mathbf{res}^{\pi}$. This is necessary in order to capture all the variability of the two-stage procedure and to correctly generate the null hypothesis of interest. Moreover, the cross-validation nature of the procedure protects against systematic bias of the residuals obtained in Step 3 based on $y^{*}$ \citep[see Chapter 7 of][]{Hastie1}.

Given the aforementioned relations between $Q_{\mathbf{X}_1}^2$, $Q_{\mathbf{X}_2|\mathbf{X}_1}^2$, and $Q_{\mathbf{X}_{1},\mathbf{X}_{2}}^2$ specified by expression (6) , note that $H_{0}:Q_{\mathbf{X}_{2}|\mathbf{X}_{1}}^2=0$ is equivalent to  $\widetilde{H}_{0}:Q_{\mathbf{X}_{1},\mathbf{X}_{2}}^2-Q_{\mathbf{X}_{1}}^2=0$. This result immediately follows from expression (6), given that $Q_{\mathbf{X}_{2}|\mathbf{X}_{1}}^2=0$ if and only if $1-Q_{\mathbf{X}_{2}|\mathbf{X}_{1}}^2=1$ (assuming $Q^{2}_{\mathbf{X}_{1}} \neq 1$). Hence, both tests are equivalent provided that the distribution under the null hypothesis is generated by the aforementioned permutation procedure, i.e.,  the p-values, resulting from using $Q_{\mathbf{X}_{2}|\mathbf{X}_{1}}^{2}$ as test statistic or  $Q_{\mathbf{X}_{1},\mathbf{X}_{2}}^{2}-Q_{\mathbf{X}_{1}}^{2}$ are approximately the same.

Due to the reliance of our method on resampling techniques, computational cost is a potential limitation of our approach. However, our method is easy to split in $M$ independent realizations of the same computational procedure. Hence, we can use parallel computing to speed up the procedure \cite{Herlihy}.

\section{Simulation study}
\subsection{Simulation setup}
We simulate two omic predictor sources $\mathbf{X}_1$ (dimension $n\times p$) and $\mathbf{X}_2$ (dimension $n \times q$) and a $n\times1$ vector  $\mathbf{y}$, the continuous outcome. We use matrix singular value decomposition (svd, \cite{Jolliffe}) of each of the two omic sources to generate common `latent' factors associated with $\mathbf{X}_1$, $\mathbf{X}_2$ and $\mathbf{y}$. Common eigenvectors in the svd of $\mathbf{X}_{1}$ and $\mathbf{X}_{2}$ introduce correlation among the omic sources. We consider different patterns in terms of the conditional association between $\mathbf{X}_{2}$ and $\mathbf{y}$ (see Figure 1). The details of the data generation procedure are as follows:


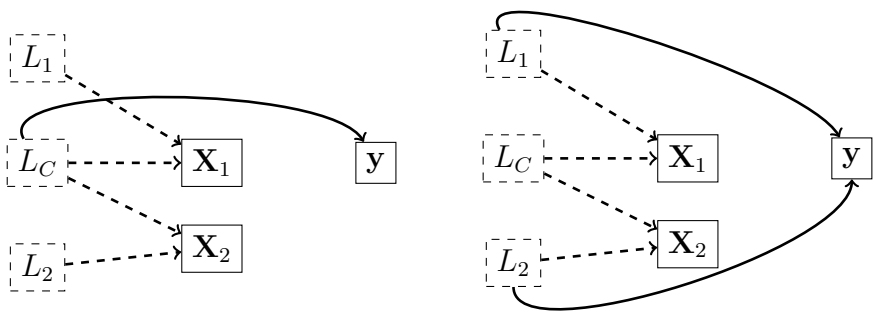
\begin{figure}[h]
\begin{minipage}{0.45\linewidth}
\begin{tikzpicture}
\node[draw,text centered] (X1) {$\mathbf{X}_{1}$};
\node[draw, below = 0.5 of X1, text centered] (X2) {$\mathbf{X}_{2}$};
\node[draw,left = 1.5 of X1, rectangle,dashed,text centered] (LJ) {$L_{C}$};
\node[draw, rectangle, dashed, above = 0.75 of LJ, text centered] (L1) {$L_{1}$};
\node[draw, rectangle, dashed, below = 0.75 of LJ, text centered] (L2) {$L_{2}$};
\node[draw,right = 1.5 of X1,text centered] (y) {$\mathbf{y}$};
 
\draw[->, line width= 1,dashed] (L1) --  (X1);
\draw [->, line width= 1,dashed] (L2) -- (X2);
\draw [->, line width= 1,dashed] (LJ) -- (X2);
\draw [->, line width= 1,dashed] (LJ) -- (X1);
\draw[->, line width=1] (LJ) to  [out=120,in=120, looseness=0.5]  (y);
\end{tikzpicture}
\end{minipage}
\begin{minipage}{0.45\linewidth}
\begin{tikzpicture}
\node[draw,text centered] (X1) {$\mathbf{X}_{1}$};
\node[draw, below = 0.5 of X1, text centered] (X2) {$\mathbf{X}_{2}$};
\node[draw,left = 1.5 of X1, rectangle,dashed,text centered] (LJ) {$L_{C}$};
\node[draw, rectangle, dashed, above = 0.75 of LJ, text centered] (L1) {$L_{1}$};
\node[draw, rectangle, dashed, below = 0.75 of LJ, text centered] (L2) {$L_{2}$};
\node[draw,right = 1.5 of X1,text centered] (y) {$\mathbf{y}$};
 
\draw[->, line width= 1,dashed] (L1) --  (X1);
\draw [->, line width= 1,dashed] (L2) -- (X2);
\draw [->, line width= 1,dashed] (LJ) -- (X2);
\draw [->, line width= 1,dashed] (LJ) -- (X1);
\draw[->, line width=1] (L1) to  [out=120,in=120, looseness=0.5]  (y);
\draw[->, line width=1] (L2) to  [out=270,in=270, looseness=0.5]  (y);
\end{tikzpicture}
\end{minipage}
\caption{Simulation study. $\mathbf{X}_{1}$ and $\mathbf{X}_{2}$ are two omic predictors sources and $\mathbf{y}$ is the outcome to be predicted. $\mathbf{L}_{1}$, $\mathbf{L}_{2}$ and $\mathbf{L}_{C}$ are three independent non-observed matrices used to generate $\mathbf{X}_{1}$ and $\mathbf{X}_{2}$. Correlation between $\mathbf{X}_{1}$ and $\mathbf{X}_{2}$ is induced by $\mathbf{L}_{C}$.  Left: Null case. No independent (of $\mathbf{X}_{1}$) association between  $\mathbf{X}_2$ and $\mathbf{y}$ ($\mathbf{y}$ generated as a linear combination of columns of $\mathbf{L}_{C}$ plus independent noise). Right: Alternative case. Independent (of $\mathbf{X}_{1}$) association between $\mathbf{X}_2$ and $\mathbf{y}$ ($\mathbf{y}$ generated as a linear combination of $\mathbf{L}_{1}$ and $\mathbf{L}_{2}$ plus independent noise).}
\end{figure}

\textbf{Step 1} Generate $\mathbf{L}\sim N(0,\mathbf{I}_{R})$,  a matrix of $r=1,\ldots,R$ i.i.d.  latent factors of $\mathbf{X}_1$ and $\mathbf{X}_2$. 

\textbf{Step 2} Define $\mathbf{\Sigma}_{1}$ ($p \times p$) and $\mathbf{\Sigma}_{2}$ ($q \times q$), the correlation matrices of $\mathbf{X}_{1}$ and $\mathbf{X}_{2}$, respectively, according to a predefined covariance structure of interest. Following the recent literature on pathway and network analysis of omics data \cite{Zhang}, we generated $\boldsymbol{\Sigma}_{i}$, $i=1,2$ according to a hub observation model \cite[][see Figure 2]{Hardin}.

\textbf{Step 3} Draw $\mathbf{X^{*}}_{i}\sim N(0,\mathbf{\Sigma}_{i})$, $i=1,2$, 
and obtain the singular value decomposition for each of the independent matrices $\mathbf{X}_{1}^{*}$ and $\mathbf{X}_{2}^{*}$:  $\mathbf{X}_{i}^{*}=\mathbf{U^{*}}_{i}\mathbf{D}_{i}\mathbf{V}_{i}^{T}, \quad i=1,2$.

\textbf{Step 4} Generate the final correlated $\mathbf{X}_{1}$ and $\mathbf{X}_{2}$ by  manipulation of $\mathbf{U^{*}}_{1}$ and $\mathbf{U^{*}}_{2}$, the left eigenvectors matrices from $\mathbf{X^{*}}_{1}$ and $\mathbf{X^{*}}_{2}$, respectively. Specifically, for a certain number ($C$) of predefined columns $C_{1}$ and $C_{2}$, the original submatrices $\mathbf{U^{*}}_{1C_{1}}$ and $\mathbf{U^{*}}_{2C_{2}}$ (independent) are replaced by $C$ common independent latent factors $L_{c}$, $c=1,\ldots,C$ generated in \textbf{Step 1}. 
In this manner, correlation between $\mathbf{X}_{1}$ and $\mathbf{X}_{2}$ is induced, while the within-omic source correlation structures $\boldsymbol{\Sigma}_{1}$ and $\boldsymbol{\Sigma}_{2}$ are preserved. 

\textbf{Step 5} Simulate the outcome $\mathbf{y}=\mathbf{X}_{1}\boldsymbol{\beta}_{1}+\mathbf{X}_{2}\boldsymbol{\beta}_{2}+\boldsymbol\epsilon$, where $\boldsymbol{\beta}_{1}$ and $\boldsymbol{\beta}_{2}$ are vectors of regression coefficients of length $p$ and $q$, respectively and $\boldsymbol\epsilon\sim N(0,1)$. Since $\mathbf{X}_{i}=\mathbf{U}_{i}\mathbf{D}_{i}\mathbf{V}_{i}^{T}$, $i=1,2$, we can rewrite $\mathbf{X}_{i}\boldsymbol{\beta}_{i}=\mathbf{U}_{i}\mathbf{D}_{i}\boldsymbol{\beta}^{*}_{i }$ and thus $\boldsymbol{\beta}_{i}=\mathbf{V}_{i}\boldsymbol{\beta}^{*}_{i}$, where $\boldsymbol{\beta}^{*}_{i}$ represents the association between $\mathbf{X}_i$ and the outcome $\mathbf{y}$ through the orthogonal directions given by $\mathbf{U}_{i}$. Consequently, we first generate $\boldsymbol\beta^{*}_{i}$ and we then transform it to the predictor space by using $\boldsymbol{\beta}_{i}=\mathbf{V}_{i}\boldsymbol{\beta}^{*}_{i}$. 

\begin{figure}[h]
\begin{minipage}{0.45\linewidth}
\includegraphics[width=5cm,height=5cm]{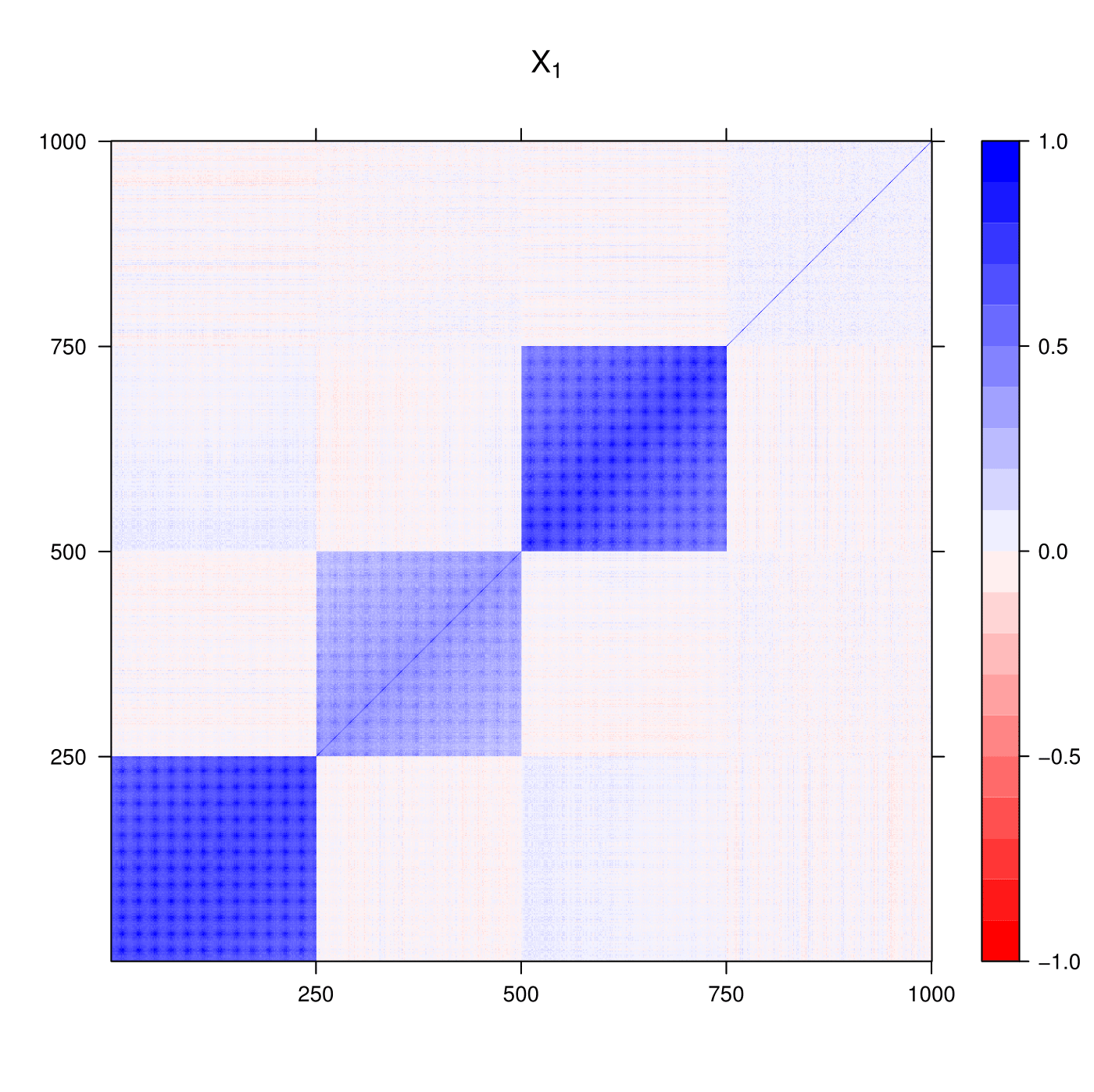}
\end{minipage}
\begin{minipage}{0.45\linewidth}
\includegraphics[width=5cm,height=5cm]{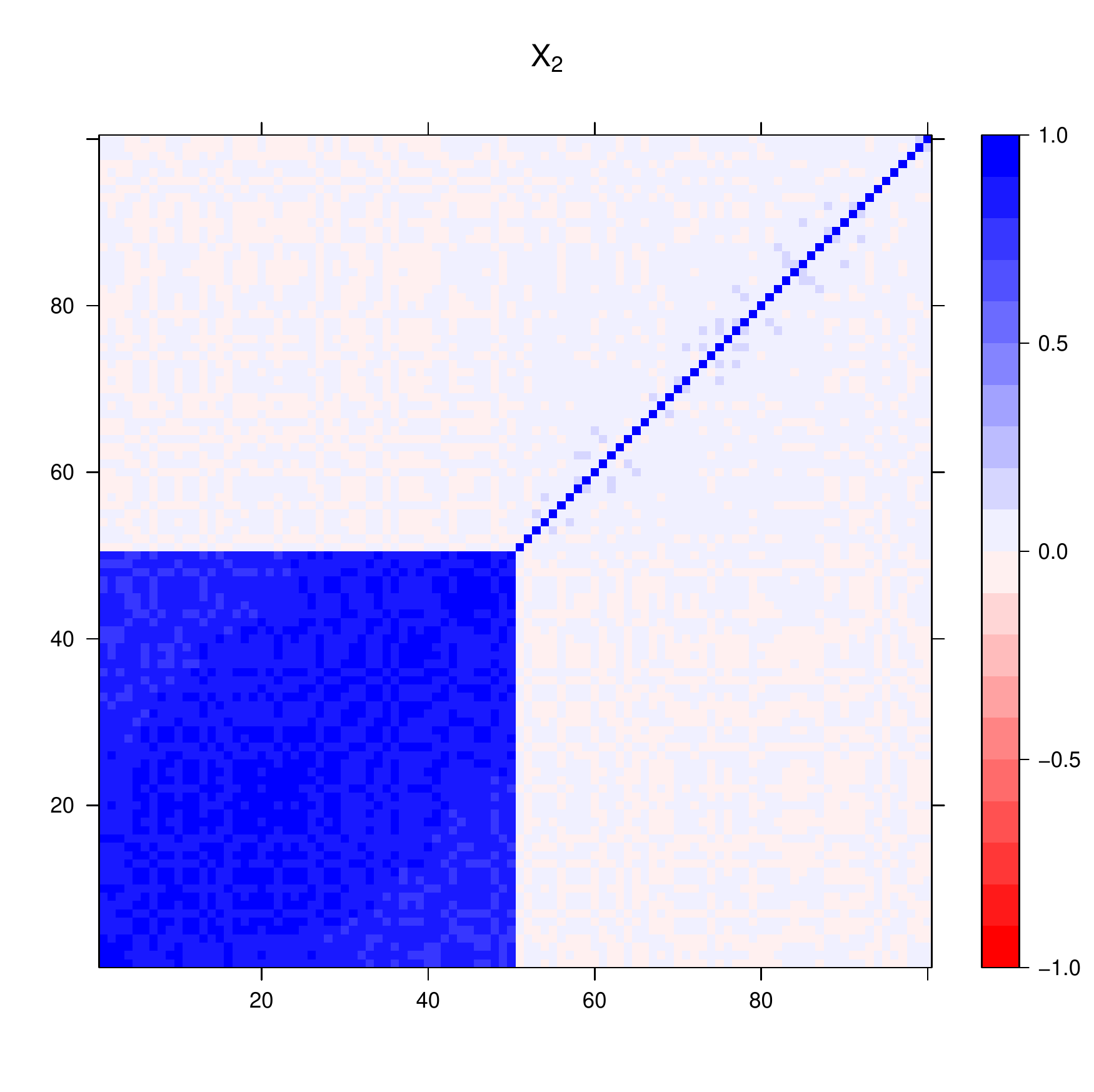}
\end{minipage}
\caption{Simulation study. Left: Correlation matrix of $\mathbf{X}_1$ ($p=1000$), 4 groups of 250 features each. Right: Correlation matrix of $\mathbf{X}_2$ ($q=100$), 2 groups of 50 features each.}
\end{figure}

\noindent\textbf{Simulation 1 (`Null' scenarios):}
The second omic $\mathbf{X}_{2}$ source is non-informative, i.e., $\boldsymbol{\beta}_{2}^{*}=\mathbf{0}$, but is strongly correlated to $\mathbf{X}_{1}$, by imposing common first columns of $\mathbf{U}_{1}$ and $\mathbf{U}_{2}$ ($\mathbf{U}_{11}=\mathbf{U}_{21}$, the correlation between omic sources is driven through the maximal variance subspace). We considered different assumptions regarding the regression dependence of $\mathbf{y}$ on $\mathbf{X}_{1}$ which has an impact on  the ability to calibrate prediction rules based on $\mathbf{X}_1$ for  $\mathbf{y}$. We consider two situations in which the association with $\mathbf{y}$ is unifactorial, in the sense that only one latent factor (one column of $\mathbf{U}_1$) is associated with $\mathbf{y}$ and two multi-factorial situations. One of our objectives is to illustrate how changing the complexity of the calibration of a prediction rule based on $\mathbf{X}_{1}$ (by formulating the problem through regression on either larger or smaller variance latent factors) may affect the results. We consider the following `null' scenarios: 
\begin{description}
\item[Scenario 1a] $\beta_{1m}^{*}=0.01$, $m=1$; $\beta_{1m}^{*}=0,\quad m \neq 1$.  $\mathbf{y}$ is associated to high-variance subspace of $\mathbf{U}_{1}$, corresponding to the largest eigenvalue of $\mathbf{X}_{1}$.

\item[Scenario 1b] $\beta_{1m}^{*}=1$, $m=6$, $\beta_{1m}^{*}=0,\quad m \neq 6$. The association with $\mathbf{y}$ relies on a low-variance subspace of $\mathbf{U}_{1}$. Hence, we expect lower values of $Q^{2}_{\mathbf{X}_1}$, compared to Scenario 1a.

\item[Scenario 1c] $\beta_{1m}^{*}=1,\quad m=1,2$, $\beta_{1m}^{*}=0$ otherwise. In this setting we consider a bifactorial regression, as association with $\mathbf{y}$ is a combination of the effect of the two first eigenvectors of $\mathbf{X}_1$. 

\item[Scenario 1d] $\beta_{1m}^{*}=1,\quad m=1,\ldots, 4$, $\beta_{1m}^{*}=0$ otherwise. In this setting we consider a multifactorial regression, as association with $\mathbf{y}$ is a combination of the effect of the four first eigenvectors of $\mathbf{X}_1$. 
\end{description}

\noindent\textbf{Simulation 2 (`Alternative' scenarios):}
$\mathbf{X}_{2}$ is associated with $\mathbf{y}$ through latent factors non-shared with $\mathbf{X}_{1}$. The following `alternative' scenarios are investigated:

\begin{description}
\item[Scenario 2a] $\boldsymbol{\beta}_{1m}^{*}=\boldsymbol{\beta}_{2m}^{*}=0.01$, $m=1$, $\boldsymbol{\beta}_{1m}^{*}=\boldsymbol{\beta}_{2m}^{*}=0$, $m\neq1$. The eigenvector related to the largest eigenvalue of each source is associated to $\mathbf{y}$ and the association between $\mathbf{X}_{1}$ and $\mathbf{X}_{2}$ is generated by sharing the second eigenvectors, i.e., by setting $\mathbf{U}_{12}=\mathbf{U}_{22}$.
\item[Scenario 2b] $\boldsymbol{\beta}_{1m}^{*}=0.01$, $m=1$, $\boldsymbol{\beta}_{1m}^{*}=0$, $m\neq1$  and $\boldsymbol{\beta}_{2m}^{*}=0.01$, $m=3$, $\boldsymbol{\beta}_{2m}^{*}=0$, $m\neq3$, and the association between $\mathbf{X}_{1}$ and $\mathbf{X}_{2}$ is generated by setting $\mathbf{U}_{11}=\mathbf{U}_{21}$.
\item[Scenario 2c] $\boldsymbol{\beta}_{1m}^{*}=0.01$, $m=6$, $\boldsymbol{\beta}_{1m}^{*}=0$, $m\neq1$  and $\boldsymbol{\beta}_{2m}^{*}=0.01$, $m=3$, $\boldsymbol{\beta}_{2m}^{*}=0$, $m\neq3$, and the association between $\mathbf{X}_{1}$ and $\mathbf{X}_{2}$ is generated by setting $\mathbf{U}_{11}=\mathbf{U}_{21}$.
\end{description}


Figure 3 shows a Monte Carlo approximation based on a sample of $n=10,000$ observations of the regression coefficients $\boldsymbol{\beta}_{1}$ and $\boldsymbol{\beta}_{2}$ in the studied simulated scenarios. From panels 3(a) to 3(d), we can observe that the different simulation settings differ in the level of imposed sparsity in the association between $\boldsymbol{y}$ and $\mathbf{X}_{1}$. On the one hand, scenarios presented in panels 3(a) (scenarios 1a, 2a ad 2b) and 3(b) (scenario 1b) are relatively sparse, with most of the simulated coefficients close to zero. On the other hand, the $\boldsymbol{\beta}_{1}$ of scenario 1c (represented in panel 3(c)) and specially of scenario 1d (represented in panel 3(d)) are less sparse, based on a large number of non-null regression coefficients in $\mathbf{X}_{1}$. With regard to $\boldsymbol{\beta}_{2}$, panel 3(e) (scenario 2a) represents a sparser situation than panel 3(f) (scenarios 2b and 2b).

\begin{figure}
\includegraphics[scale=0.22]{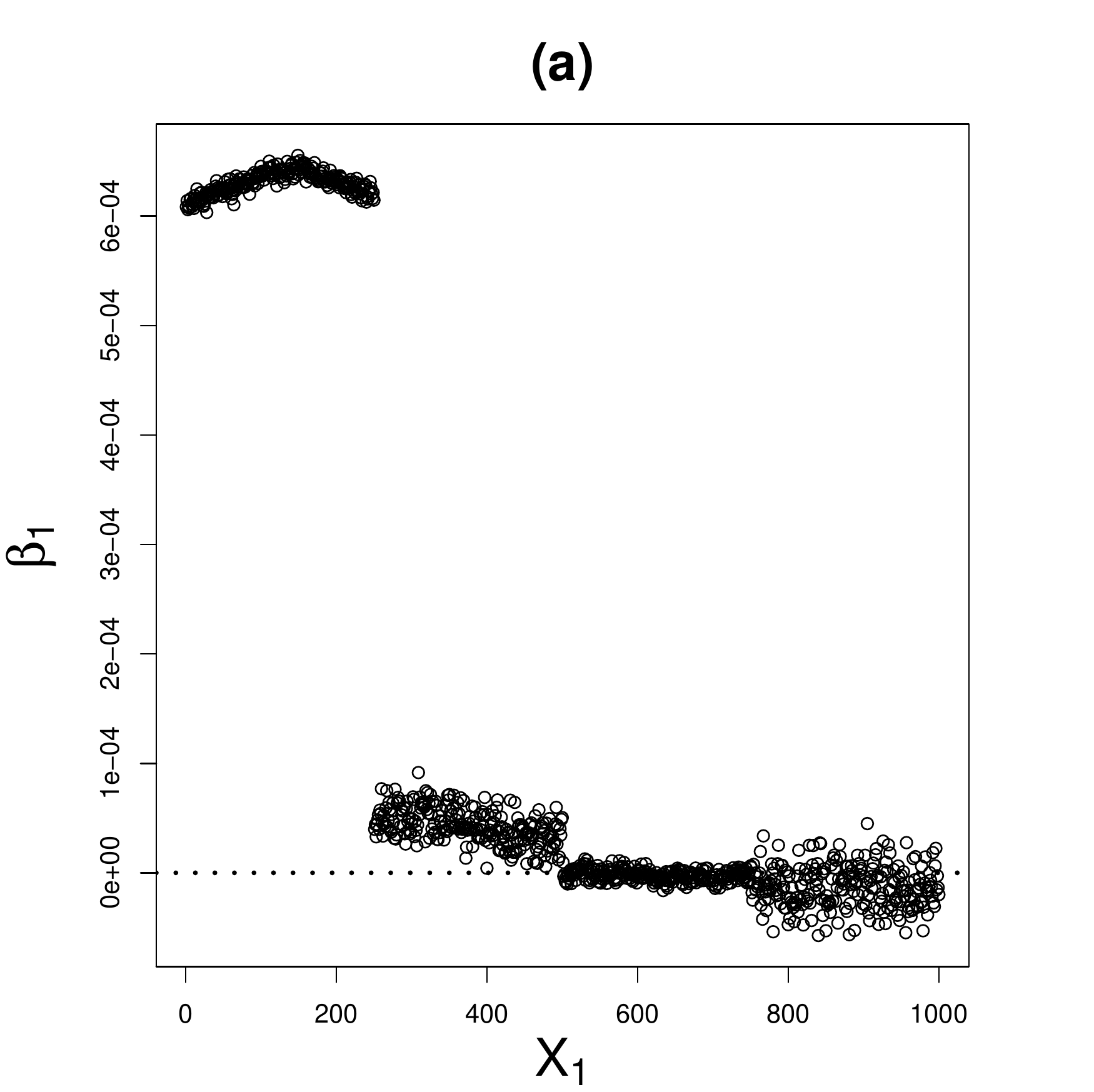}
\includegraphics[scale=0.22]{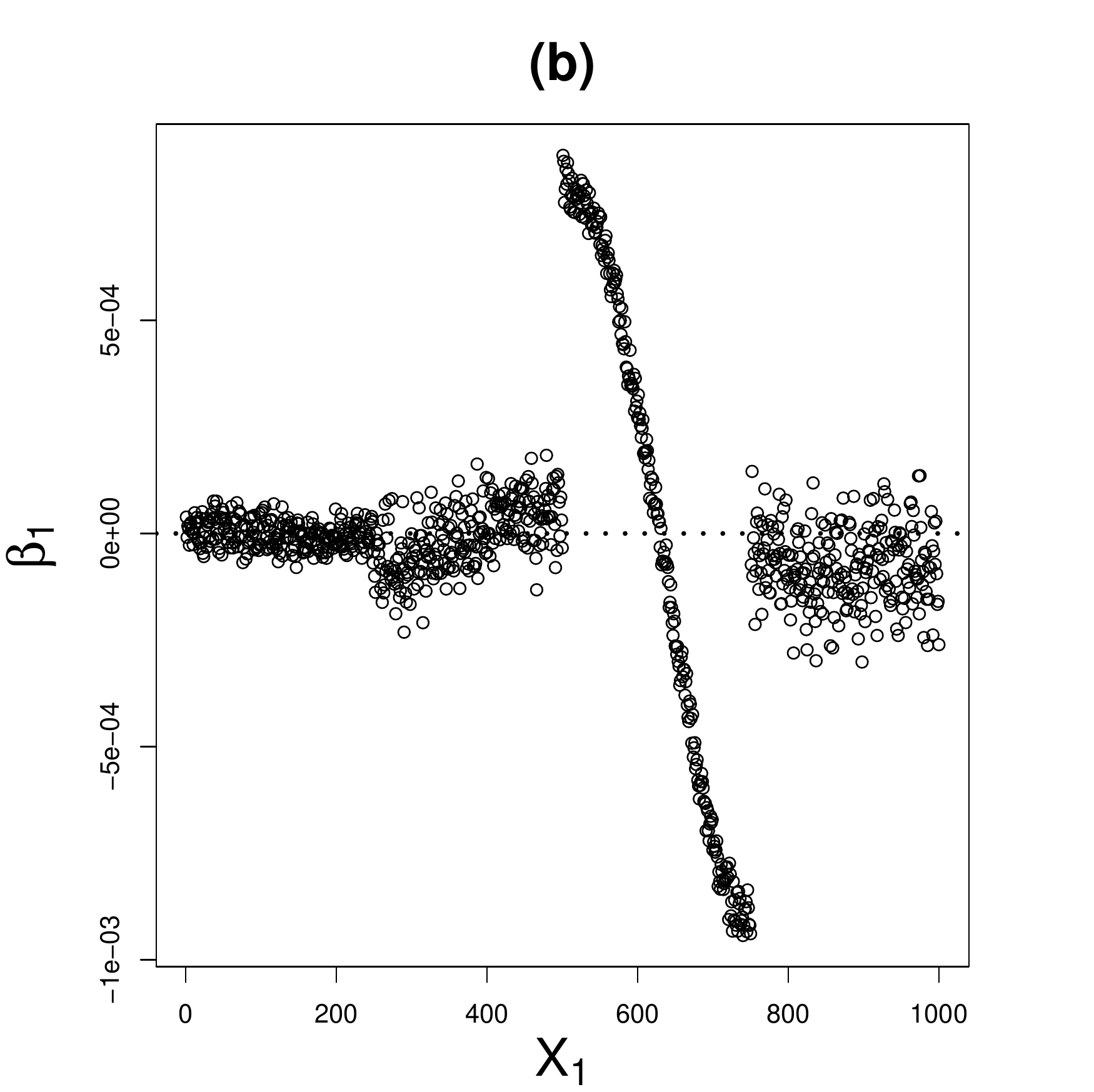}
\includegraphics[scale=0.22]{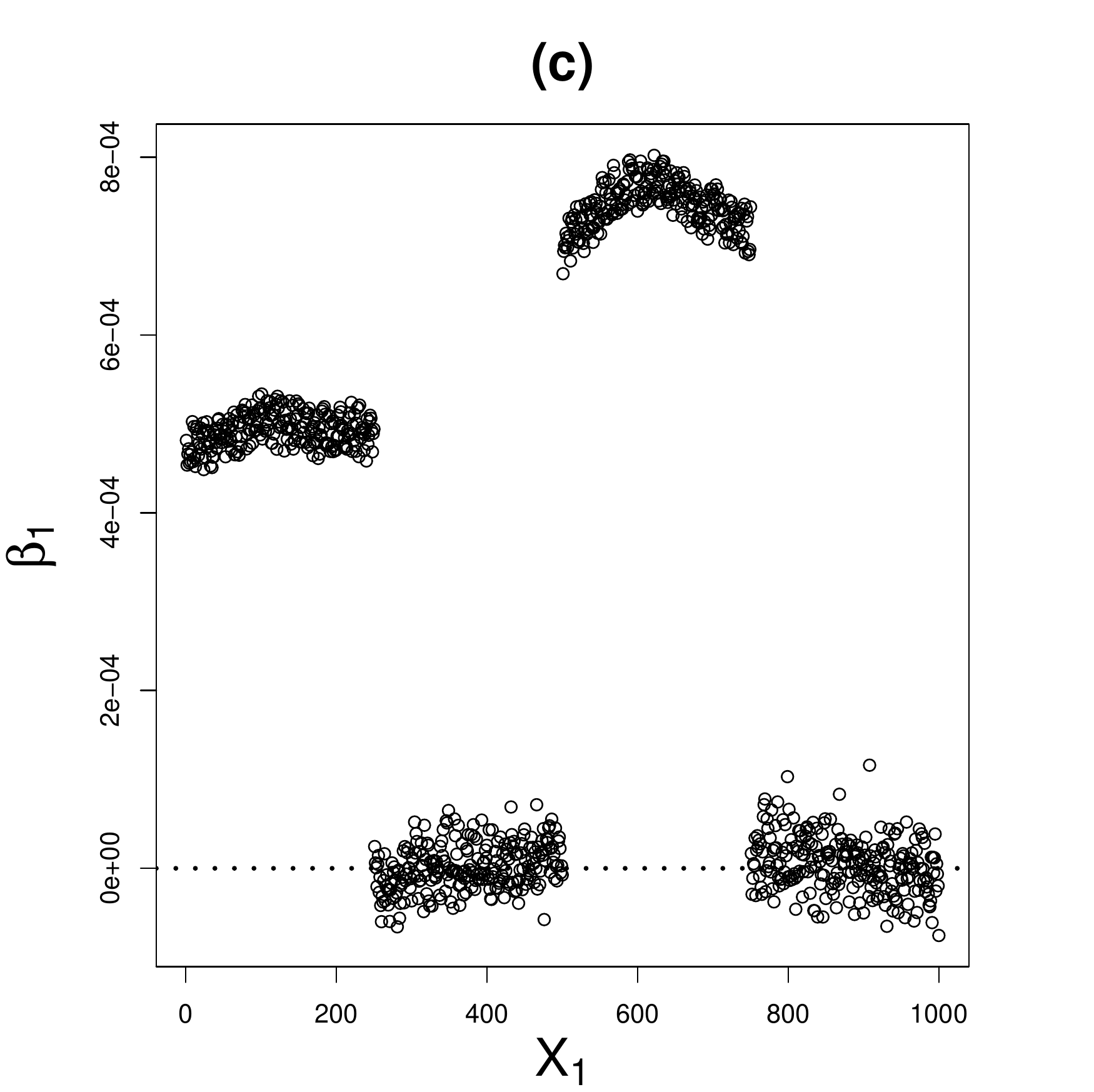}\\
\includegraphics[scale=0.22]{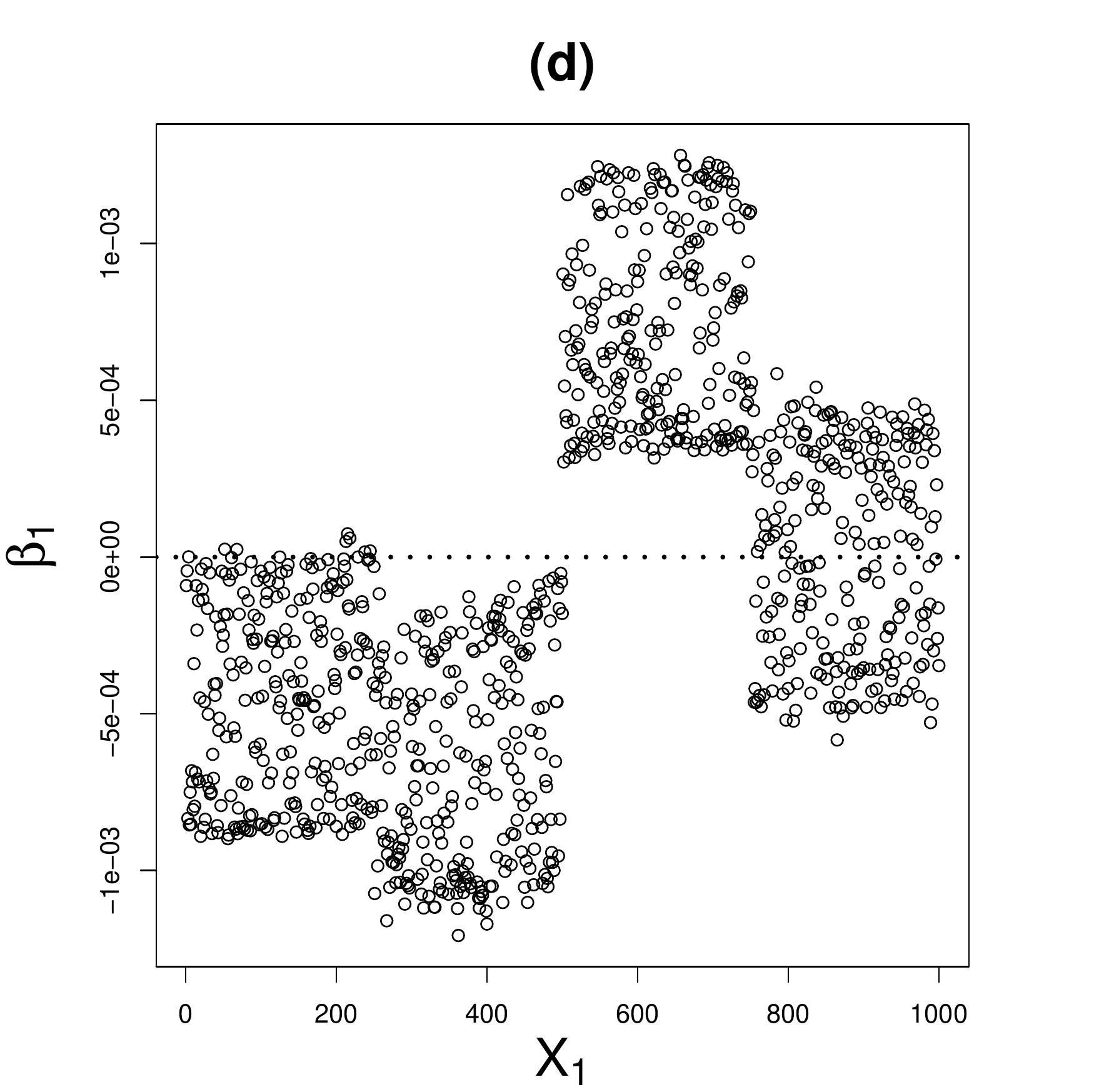}
\includegraphics[scale=0.22]{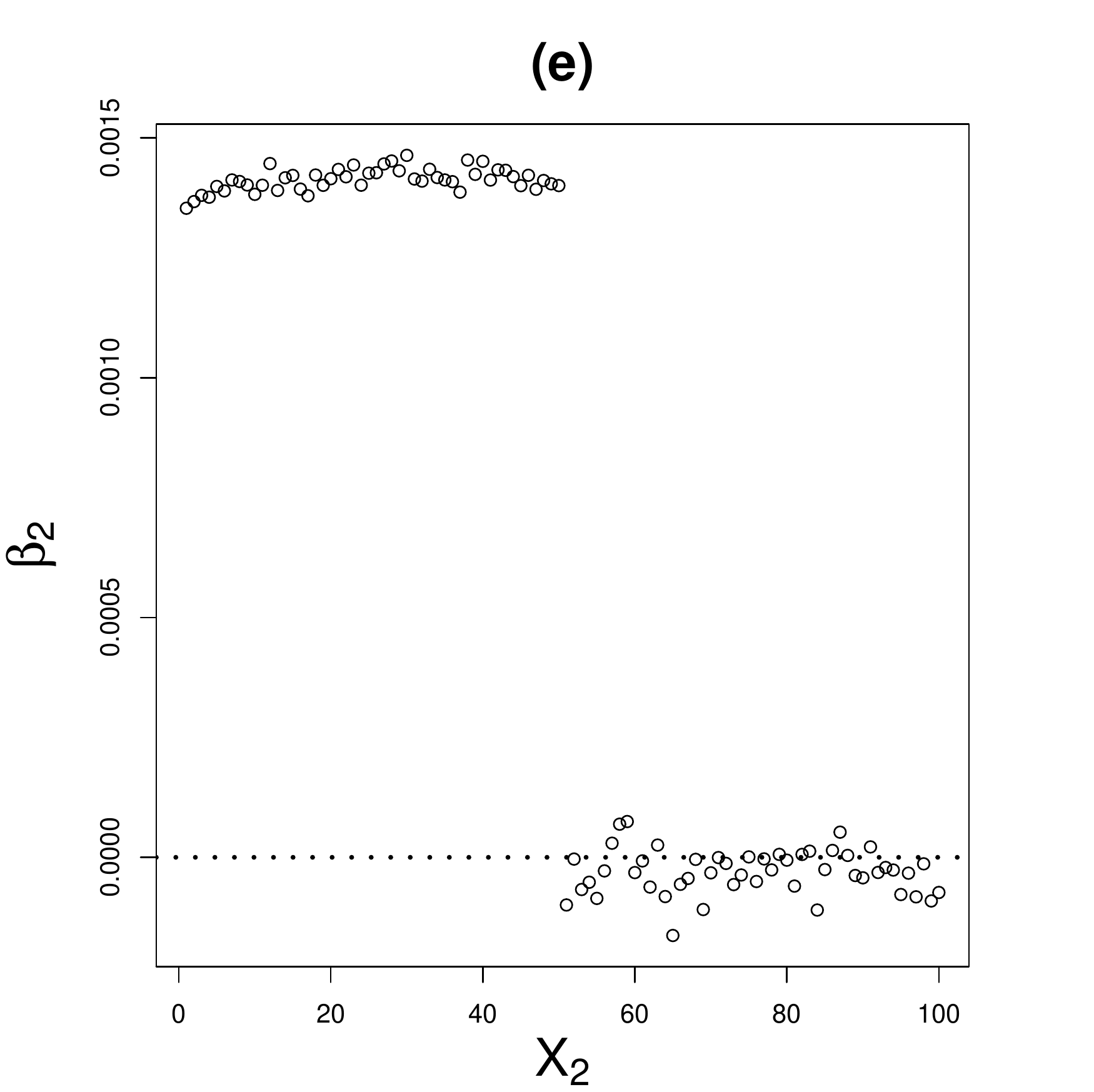}
\includegraphics[scale=0.22]{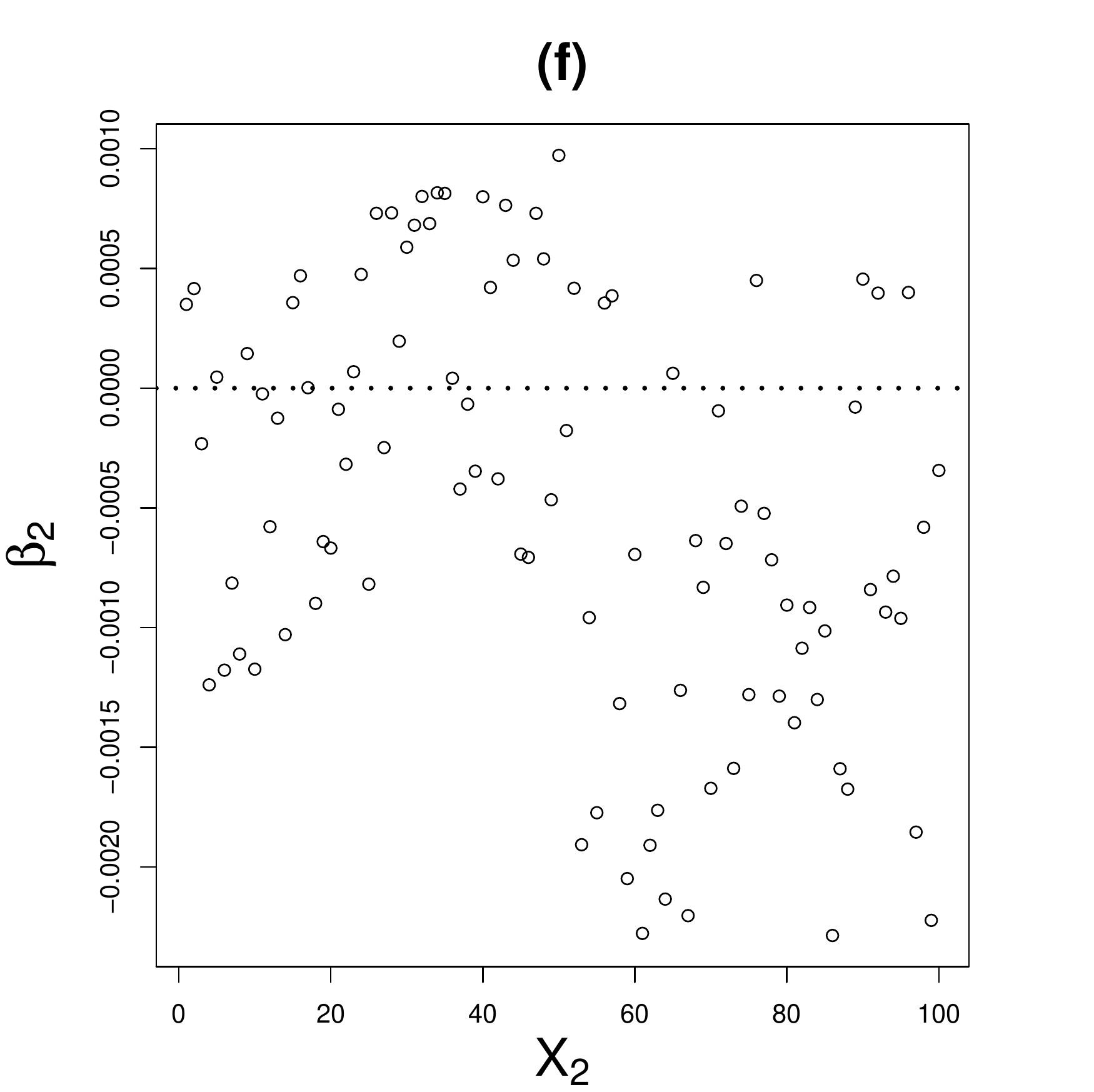}
\caption{Simulation study. (a)-(d): regression coefficients (elements of $\boldsymbol{\beta}_{1}$, y-axis) corresponding to each of the $p$ predictors of $\mathbf{X}_{1}$ (x-axis). (c)-(d): regression coefficients (elements of $\boldsymbol{\beta}_{2}$, y-axis) corresponding to each of the $q$ predictors of $\mathbf{X}_{2}$ (x-axis). The outcome variable is generated as  $\mathbf{y}=\mathbf{X}_{1}\boldsymbol{\beta}_{1}+\mathbf{X}_{2}\boldsymbol{\beta}_{2}+\boldsymbol\epsilon$, $\boldsymbol\epsilon\sim N(0,1)$.(e)-(f) provide information about association between $\mathbf{y}$ and $\mathbf{X}_1$ and (e) and (f) corresponds to the independent association between $\mathbf{y}$ and $\mathbf{X}_2$ in the alternative scenarios (for the null scenarios 1a-1d the independent association between $\mathbf{y}$ and $\mathbf{X}_2$ is null). (a) corresponds to scenarios 1a, 2a and 2b, (b) correspond to scenarios 1b and 2c respectively, while (c) and (d) correspond to scenario 1c and 1d. (e) shows the association ($\boldsymbol{\beta}_{2}$) between $\mathbf{X}_2$ and $\mathbf{y}$ in scenario 2a and (e) shows $\boldsymbol{\beta}_{2}$ for scenarios 2b and 2c.}
\end{figure}

In our basic setting, we considered $n=100$ observations, $p=1000$ features in $\mathbf{X}_{1}$ and $q=100$ features in $\mathbf{X}_{2}$. For each scenario, we provide the mean values and standard deviations of $Q_{\mathbf{X}_{1}}^2$, $Q_{\mathbf{X}_{2}|\mathbf{X}_{1}}^2$, and $Q_{\mathbf{X}_{1},\mathbf{X}_{2}}^2$, based on 5-folds double cross-validation, jointly with the rejection proportions for testing $H_{0}:Q_{\mathbf{X}_{2}|\mathbf{X}_{1}}^{2}$ along $M=500$ Monte Carlo trials. We evaluated the permutation test introduced in Subsection 2.3. using $n_{perm}=200$ permutations. We complemented our empirical evaluations of the proposed sequential double cross-validation procedure by extending our basic simulation setting in two directions. We checked the impact on modifying sample size ($n=50$) and the complexity of the problem by varying the number of variables considered in the first stage ($p=4000$).


Additionally, we compared the performance  of our procedure based on double-cross validation with two alternative strategies. On the one hand, we provide results based on a two-stage procedure using a single cross-validation loop (cross-validation is used for model choice but predictions and therefore the residuals used as outcome in the second stage are directly computed on the complete sample). On the other hand, we check the impact on the results of over-penalization. Specifically, instead of taking $\lambda_{opt}$ as defined in the inner loop of the double cross-validation procedure presented in Subsection 2.1., we choose a larger value for $\lambda$, namely $\lambda_{opt}+1 s.e.(\lambda_{opt})$. Both are usual strategies in penalized regression in single omic prediction frameworks, so it is of practical interest to quantify their impact from the added predictive value point of view. The results of these alternative strategies are provided as Supplemental material but discussed in the main text.

\subsection{Simulation results}

The results for the sequential double cross-validation procedure (labeled as `CV type= $CV_{D}$, $\lambda_{opt}$') are summarized in Tables 1 and 2. The top part of each table contains the results concerning the `null' scenarios (no added value of $\mathbf{X}_{2}$), while the bottom part shows the results of the `alternative' scenarios (added value of $\mathbf{X}_{2}$). Table 1 contains results based on ridge regression, $\alpha=0$ in expression (1), while Table 2 summarizes the results for the lasso penalty type ($\alpha=1$).

\subsubsection{Ridge regression}

For the four `null' scenarios 1a-1d, given that $\mathbf{X}_{2}$ is not independently associated to $\mathbf{y}$, we expect $Q_{\mathbf{X}_{2}|\mathbf{X}_{1}}^{2}=0$ and rejection proportions of $H_{0}$ about 0.05. 

The results of the sequential double cross-validation procedure based on ridge regression are satisfactory in this regard, with rejection proportions close to the nominal level in all the studied null scenarios and for different sample sizes ($n=50$, $n=100$) and levels of complexity of the first step ($p=1000$, $p=4000$).

The top part of Table 1 shows that the estimated $Q_{\mathbf{X}_{1}}^{2}$ for scenarios 1a, 1c and 1d are large and very similar ($Q_{\mathbf{X}_{1}}^{2}\sim 0.90$). As it was expected, the estimated predictive ability of $\mathbf{X}_{1}$ is lower in scenario 1b and presents a larger variability, since the association between $\mathbf{y}$ and $\mathbf{X}_{1}$ relies on a small variance subspace. In general, for all 1a-1d scenarios the estimated $Q_{\mathbf{X}_{2}|\mathbf{X}_{1}}^{2}$ is close to zero. However, we observe that the sample size influences the estimated  $Q_{\mathbf{X}_{1}}^2$ and hence, due to the correlation between $\mathbf{X}_{1}$ and $\mathbf{X}_{2}$, also affects the estimation of $Q_{\mathbf{X}_{2}|\mathbf{X}_{1}}^{2}$. We observe systematically lower values of $Q_{\mathbf{X}_{1}}$ for $n=50$ than for $n=100$ in all the studied `null' scenarios. This feature translates in systematically larger values of $Q_{\mathbf{X}_{2}|\mathbf{X}_{1}}^{2}$ for $n=50$ than for $n=100$. However, the permutation test is able to account for this issue and the level of the test is respected independently of the sample size. Analogously, increasing the number of features of the first source $\mathbf{X}_{1}$ (from $p=1000$ to $p=4000$) while keeping fixed the number of features of $\mathbf{X}_{2}$ ($q=100$) also affects the estimation of $Q_{\mathbf{X}_{1}}$ and $Q_{\mathbf{X}_{2}|\mathbf{X}_{1}}^{2}$. In this case, the values of $Q_{\mathbf{X}_{1}}$ are larger and hence, the values of $Q_{\mathbf{X}_{2}|\mathbf{X}_{1}}^{2}$ tend to be closer to zero. Worth noting is that the level of the test is also well respected in this case.

The bottom part of Table 1 shows the results for the alternative scenarios. As desirable, the power increases with sample size for all the three studied alternative scenarios. As it was the case for the `null' scenarios, increasing the sample size tends to lead to better predictive ability of $\mathbf{X}_{1}$. This result matches intuition, since larger sample sizes provide more information for model building, and hence, under correct model specification, the resulting predicting models are expected to behave better in new data. An exception to this is scenario 2c, where our double cross-validation procedure seems to overfit with $n=50$. This is due to the fact that scenario 2c, unlike scenarios 2a and 2b, is characterized as a `difficult' prediction problem when considering $\mathbf{X}_{1}$ (association with $\mathbf{y}$ is driven by a low-variance subspace of $\mathbf{X}_{1}$). In line with this, the power of the test is different for the three different studied scenarios. The greatest power is reached in scenario 2a, in which the independent association between $\mathbf{X}_{2}$ and $\mathbf{y}$ is driven through the subspace of maximum variation and the first step of the procedure relies on a relatively `easy' prediction problem.

Even if scenarios 2b and 2c are based on the same independent association between $\mathbf{X}_2$ and $\mathbf{y}$, the impact of the first source on the power of the test is large. Scenario 2b, in which $Q_{\mathbf{X}_{1}}^{2}=0.87$ for $n=100$ reaches a power of 71 \%, while the rejection rate reduces to 19\% in scenario 2c, corresponding to a more `difficult' prediction problem in the first stage, reflected in a low and unstable $Q_{\mathbf{X}_{1}}^{2}$ ($Q_{\mathbf{X}_{1}}^{2}=0.28$ for $n=50$ and $Q_{\mathbf{X}_{1}}^{2}=0.09$ for $n=100$).

\begin{table}[htbp]
\label{Table1}
  \centering
  {\footnotesize
  \caption{Ridge ($\alpha=0$). Mean estimates (and standard deviation in brackets) of $Q_{\mathbf{X}_{1}}^2$, $Q_{\mathbf{X}_{2}|\mathbf{X}_{1}}^2$, $Q_{\mathbf{X}_{1},\mathbf{X}_{2}}^2$  and rejection proportions of the permutation test based on  $Q_{\mathbf{X}_{2}|\mathbf{X}_{1}}^{2}$ along 500 Monte Carlo trials.}
 \begin{tabular}{cc|c|c|c|c}
    \\
    
      \hline
    {\footnotesize Scenario} & $n$&$Q_{\mathbf{X}_{1}}^2$ (Step 1)    & $Q_{\mathbf{X}_{2}|\mathbf{X}_{1}}^2$ (Step 2)& $Q_{\mathbf{X}_{1},\mathbf{X}_{2}}^ 2$ (Global)&{Rej. Prop.} \\   
          \hline 
            & & &&&\\
                       &$n=50$ &0.85 (0.03)  &0.07 (0.09)&0.89 (0.05) &0.058\\     
            1a  &$n=100$ & 0.88 (0.02) &0.03 (0.04)& 0.91 (0.03)&0.068\\     
             &$n=100$, $p=4000$ &0.94 (0.02)  &0.03 (0.04)&0.99 (0.01) &0.056\\ 
          
    & & &&&\\
   &$n=50$ &0.31 (0.07) &0.04 (0.07)&0.34 (0.09) &0.044\\ 
              1b  &$n=100$& 0.41 (0.07)  &0.01 (0.02)&0.42 (0.07) &0.047\\     
           & $n=100$, $p=4000$&0.50 (0.09) &0.01 (0.03)&0.72 (0.13) &0.060\\  
          
    & & &&&\\
   
     &$n=50$ &0.86 (0.03) &0.06 (0.08)&0.86 (0.04) &0.060\\
                1c&$n=100$ &0.91 (0.02) &0.02 (0.03)&0.92 (0.02)&0.050\\     
          &$n=100$, $p=4000$ &0.92 (0.05) &0.02 (0.04)&0.97 (0.05) &0.064\\     
           
  & & &&&\\
                   &$n=50$&0.83 (0.03)&0.05 (0.08)&0.84 (0.04)&0.046\\
                   1d&$n=100$ & 0.86 (0.02)  &0.00 (0.00)&0.97 (0.01)&0.062\\     
           &$n=100$, $p=4000$ &0.88 (0.05)&0.00 (0.00)&0.97 (0.05)&0.044\\ 
   & & &&&\\ 
  \hline
  & & &&&\\   
 &$n=50$ & 0.64 (0.13) &0.50 (0.15)&0.76 (0.16) &0.936\\
             2a &$n=100$ & 0.68 (0.10)   &0.60 (0.12) & 0.91 (0.11)&0.997\\     
            &$n=100$, $p=4000$ &  0.89 (0.05)&0.59 (0.08)&0.93 (0.06) &0.996\\         
  & & &&&\\ 
  
   &$n=50$ &0.84 (0.04)  &0.16 (0.11)&0.93 (0.04) &0.236\\                            
               2b &$n=100$ &0.87 (0.03)&0.11 (0.06)&0.95 (0.01) &0.712\\     
          & $n=100$, $p=4000$& 0.88 (0.02)& 0.10 (0.06)& 0.94 (0.03) &0.652\\  
  & & &&&\\ 
     &$n=50$ &0.28 (0.07)  &0.11 (0.11)&0.36 (0.11) &0.184\\          
2c &$n=100$ &   0.09 (0.08)  &0.01 (0.00) &0.13 (0.16) &0.186\\     
    &$n=100$, $p=4000$ &0.16 (0.09)&0.08 (0.06)&0.52 (0.09) &0.526\\ 
  & & &&&\\          
  \hline              
    \end{tabular}%
  \label{tab:addlabel}%
}\end{table}%


\subsubsection{Lasso regression}
Table 2 shows the results for double cross-validation procedure based on the lasso specification ($\alpha=1$). With regard to the `null' scenarios, we observe a good performance for scenarios 1a and 1b, with rejection proportions close to the nominal level. Interestingly, the rejection proportion of the permutation test increases with sample size and the number of features in the first source in scenarios 1c and 1d, which indicates a bad performance of the procedure based on laso regression in these settings. Namely, the bad performance of the lasso specification for scenario 1d does not improve by increasing sample size (7\% of rejections with $n=50$, 9\% of rejections with $n=100$ and 36\% of rejections for $p=4000$). The reason behind this difference with the ridge-based results is the mis-specification of the lasso with respect to the underlying data-generating mechanism. Lasso regression assumes that the true model is sparse, while,  as mentioned, scenario 1c and specially 1d correspond to non-sparse solutions.

These findings illustrate how model mis-specification may result in an improvement of predictions by adding a second source of predictors, not because of independent association to the outcome, but just because of the correlation with the first source of predictors.  

The bottom part of Table 1 shows the results for the alternative scenarios.
With respect to the alternative scenarios (bottom part of Table 2), the conclusions are similar to those observed  for ridge regression. The power increases with the sample size, and the rejection proportions differ across the three scenarios. However, we observe that ridge outperforms lasso in terms of power, specially for scenarios 2a and 2b.

\begin{table}[htbp]
  \centering
  {\footnotesize
  \caption{Lasso ($\alpha=1$). Mean estimates (and standard deviation in brackets) of $Q_{\mathbf{X}_{1}}^2$, $Q_{\mathbf{X}_{2}|\mathbf{X}_{1}}^2$, $Q_{\mathbf{X}_{1},\mathbf{X}_{2}}^2$  and rejection proportions of the permutation test based on  $Q_{\mathbf{X}_{2}|\mathbf{X}_{1}}^{2}$ along 500 Monte Carlo trials.}
    \begin{tabular}{cc|c|c|c|c}
    \\
    
      \hline
    {\footnotesize Scenario} & $n$&$Q_{\mathbf{X}_{1}}^2$ (Step 1)    & $Q_{\mathbf{X}_{2}|\mathbf{X}_{1}}^2$ (Step 2)& $Q_{\mathbf{X}_{1},\mathbf{X}_{2}}^ 2$ (Global)&{Rej. Prop.} \\
          \hline 
            & & &&&\\   
             & $n=50$&0.79 (0.06)  &0.15 (0.14)&0.88 (0.05) &0.058\\ 
             1a &$n=100$ &  0.86 (0.03)  &0.05 (0.06)&0.91 (0.03)&0.054\\     
           & $n=100$, $p=4000$&0.89 (0.02)  &0.10 (0.06)&0.95 (0.01) &0.140\\
  & & &&&\\ 
   &$n=50$ &0.18 (0.11) &0.08 (0.12)&0.27 (0.16) &0.056\\
              1b  &$n=100$ & 0.33 (0.09)  &0.02 (0.04)&0.35 (0.09) &0.056\\     
           &$n=100$, $p=4000$ &0.44(0.07) &0.03 (0.05)&0.45 (0.07) &0.058\\
   & & &&&\\
     & $n=50$&0.80 (0.05) &0.12 (0.13)&0.84 (0.05) &0.054\\ 
                1c&$n=100$ & 0.89 (0.02)  &0.04 (0.05)&0.90 (0.03) &0.068\\     
             &$n=100$, $p=4000$&0.85 (0.28) &0.13 (0.09)&0.90 (0.03) &0.352\\ 
  & & &&&\\ 
   &$n=50$&0.67 (0.06)&0.14 (0.12)&0.72 (0.07)&0.070\\ 
                1d   &$n=100$ & 0.83 (0.03)   &0.05 (0.05)&0.85 (0.03)&0.094\\     
           &$n=100$, $p=4000$ &0.73 (0.04)&0.12 (0.09)&0.78 (0.04)&0.360\\ 
          & & &&&\\   
  \hline
    & & &&&\\ 
       & $n=50$& 0.58 (0.09) &0.44 (0.14)&0.69 (0.13) &0.768\\         
             2a  &$n=100$ &0.67 (0.07)   &0.53 (0.09)&0.90 (0.03)&1.000\\     
           & $n=100$, $p=4000$&0.84 (0.04)  &0.47 (0.09)& 0.84 (0.06)&1.000\\         
    & & &&&\\  
     & $n=50$&0.77 (0.07) &0.21 (0.15) &0.89 (0.06)  &0.106\\  
               2b &$n=100$ & 0.84 (0.04)  &0.12 (0.09)&0.92 (0.02) &0.481\\     
         &$n=100$, $p=4000$ &0.83 (0.02) &0.18 (0.09) &0.96 (0.02) &0.506\\ 
    & & &&&\\ 
     & $n=50$&0.15 (0.10)&0.12 (0.13)&0.28 (0.17) &0.086\\
                2c &$n=100$ &0.27 (0.09)   &0.09 (0.07) &0.33 (0.11)&0.351\\     
    &$n=100$, $p=4000$&0.07 (0.07)&0.06 (0.06)&0.43 (0.07) &0.227\\     
     & & &&&\\          
  \hline       
    \end{tabular}%
  \label{tab:addlabel}%
}\end{table}%

\subsubsection{Alternative procedures}
Tables S3 and S4 summarize the results for the two aforementioned alternative strategies in the basic setting ($p=1000$, $q=100$) and two sample sizes ($n=50$, $n=100$):  `$CV_{D}$, $\lambda_{1se}$' corresponds to the strategy in which the sequential double cross-validation is over-shrunk (by taking $\lambda_{opt}+1s.e.(\lambda_{opt})$ instead of $\lambda_{opt}$ in the inner cross-validation loop) and `$CV_{S}$, $\lambda_{opt}$' represents the sequential procedure based on one single cross-validation loop (standard residuals as opposed to deletion-based residuals). 

In general, these two alternative strategies provide different estimates for the predictive ability of the two studied sources of predictors. Taking the double-cross validation approach  as gold-standard, we observe that the over-shrinkage of the predictions in the first step of the `$CV_{D}$, $\lambda_{1se}$' method provokes an under-estimation of $Q_{\mathbf{X}_{1}}^{2}$, while the `$CV_{S}$, $\lambda_{opt}$' provides an over-estimation, specially when the association between outcome and first source of predictors is driven through a low-variance space. For example, in the scenario 1b, for $n=100$, $Q_{\mathbf{X}_{1}}^{2}=0.69$ when based on a single cross-validation approach, notably larger than $Q_{\mathbf{X}_{1}}^{2}=0.41$ estimated by the double cross-validation approach.   Moreover, we observe that the effect of re-using the data is larger for small sample sizes, with systematically larger $Q_{\mathbf{X}_{1}}^{2}$ for $n=50$ than for $n=100$. However, under the null hypothesis, the introduced bias on the first step for both alternatives does not translate in an inflated type I error. The method labeled as `$CV_{D}$, $\lambda_{1se}$', based on double cross-validation but based on under-fitting by over-penalization controls the false discovery rate under the null hypothesis in similar fashion than the procedure introduce in Subsection 2.1. With regard to the method based on single cross-validation (`$CV_{S}$, $\lambda_{opt}$'), its behavior is slightly conservative under the null hypothesis.

For the alternative scenarios, as $Q_{\mathbf{X}_{1}}^2$, $Q_{\mathbf{X}_{2}|\mathbf{X}_{1}}^{2}$ and $Q_{\mathbf{X}_{1},\mathbf{X}_{2}}^{2}$, are underestimated by `$CV_{D}$, $\lambda_{1se}$', while `$CV_{S}$, $\lambda_{opt}$' overfits both. Even if power increases with sample size, both methods are systematically less powerful than our proposal,`$CV_{D}$, $\lambda_{opt}$, which makes it the preferable method from both an estimation and testing point of view.

\section{Application: DILGOM data}
To illustrate the performance of the proposed sequential double cross-validation procedure, and to compare it to the alternative strategies discussed in Section 3, we analyzed data from the DILGOM study. We are interested in the ability of serum NMR metabolites and microarray gene expression levels in blood to predict body mass index (BMI) at 7 years of follow-up. The metabolomic predictor data consists of quantitative information on 137 metabolic measures, mainly composed of measures on different lipid subclasses, but also amino acids, and creatine. The gene expression profiles were derived from Illumina 610-Quad SNParrays (Illumina Inc., San Diego, CA, USA). Initially, 35,419 expression probes were available after quality filtering. In addition to the pre-processing steps described by \cite{Inouye1}, we conducted a prior filtering approach and removed from our analyses probes with extremely low variation (see \cite{Liu} for details on the conducted pre-processing). As a result, we retained measures from 7380 beads for our analyses. The analyzed sample contained $n=248$ individuals for which both types of omic measurements and the BMI after 7 years of follow-up (mean=26 $kg/m^{2}$, sd=5 $kg/m^{2}$) were available. We carried out two distinct analyses using the added predictive value assessment approach described in this paper. As a first analysis, we consider the metabolic profile as primary omic source for the prediction of the log-transformed BMI and we evaluated the added predictive value of blood transcriptomics profiles. This approach is the most relevant in practice, because of both biological and economical reasons. On the one hand, metabolome (which contains, among other, cholesterol measures) is presumably more predictive of BMI than gene expression in blood. On the other hand, NMR technology is typically more affordable \cite{Soininen} than available technologies for transcriptomic profiling, so favoring the NMR source seems a sensible approach in our setting. Nevertheless, to illustrate the properties of our method, we also consider a second analysis in which we reversed the roles of the omic sources,  first fitting a model based on gene expression and then evaluating  the added predictive value of the metabolome. As in the simulation study, we considered ridge and lasso regression as prediction models, using the same alternative strategies to the sequential double cross-validation procedure presented in section 3 (`$CV_{D}$, $\lambda_{opt}$'): `$CV_{D}$, $\lambda_{1se}$', and `$CV_{S}$, $\lambda_{opt}$'. The main findings are summarized in Table 3. To check stability of the results, we artificially reduced the sample size of the available DILGOM data and checked the impact on the estimation of the added predictive ability with our sequential double cross-validation approach and its corresponding p-value. We also compared our method with a naive approach, consisting in stacking both metabolites and transcriptomics, and hence ignoring their different origin. The results of these two additional analyses are given as Supplemental Materials.

\begin{table}[htbp]
  \centering
  \caption{Application to DILGOM data. Alternative cross-validation strategies. P-values based on 1000 permutations}
    \begin{tabular}{cc|c|c|c|c}
    
      \hline
      
 $\alpha$&CV type&$Q_{Metab}^2$ & $Q_{GE|Metab}^2$&$Q_{Metab,GE}^2$&p-value \\
          \hline
          & &&   &  &\\  
$\alpha=0$&$CV_{D}$,$\lambda_{opt}$ & 0.305& 0.071  & 0.415 &$<0.001$\\
$\alpha=0$&$CV_{D}$,$\lambda_{1se}$ &0.090& 0.006  &0.102 &$<0.001$\\    
$\alpha=0$&$CV_{S}$,$\lambda_{opt}$ & 0.291& 0.137  & 0.447 &0.001\\    
 & &&   &  &\\
$\alpha=1$&$CV_{D}$,$\lambda_{opt}$ & 0.343& 0.090  & 0.458 &0.004\\
$\alpha=1$&$CV_{D}$,$\lambda_{1se}$ & 0.073&0.004  & 0.083&0.018\\    
$\alpha=1$&$CV_{S}$,$\lambda_{opt}$ &0.374 &0.060   & 0.457&0.035\\    
 & &&   &  &\\
   \hline
 $\alpha$&CV type&$Q_{GE}^2$ & $Q_{Metab|GE}^2$&$Q_{GE,Metab}^2$&p-value \\ 
           \hline
          & &&   &  &\\   
$\alpha=0$&$CV_{D}$,$\lambda_{opt}$ & 0.092& 0.194  & 0.327&$<0.001$\\ 
   $\alpha=0$&$CV_{D}$,$\lambda_{1se}$ &0.019& 0.091  & 0.117 &$<0.001$\\    
$\alpha=0$&$CV_{S}$,$\lambda_{opt}$&0.153 &0.191 & 0.380  &0.003 \\    
 & &&   &  &\\
$\alpha=1$&$CV_{D}$,$\lambda_{opt}$ & 0.277& 0.102  & 0.453&$<0.001$\\ 
$\alpha=1$&$CV_{D}$,$\lambda_{1se}$ &0.043 &0.122  &0.204  &$<0.001$\\    
$\alpha=1$&$CV_{S}$,$\lambda_{opt}$ &0.372 &0.012  & 0.409 &0.490\\ 
  \hline
    \end{tabular}%
  \label{tab:addlabel}%
\end{table}%

The upper part of Table 3 shows the results from our primary analysis, focused on evaluating the added predictive ability of gene expression with respect to our primary omic set, NMR-metabolomics, in the context of prediction of future BMI. We observe that NMR-metabolomics itself presents a moderately large predictive ability. Both ridge and lasso regression provide very similar $Q_{Metab}^{2}$ values which are slightly larger than 30\% ($Q_{Metab}^{2}=0.305$ for ridge regression, $Q_{Metab}^{2}=0.343$ for lasso regression). According to the results shown in the third and forth columns of Table 3, we observe a highly significant positive added predictive ability of gene expression with regard to log-transformed BMI after 7 years of follow-up. The size of such added predictive ability (summarized by $Q_{GE|Metab}^{2}$) indicates a modest incremental contribution of gene expression, very similar for the two considered regularization methods ($Q_{GE|Metab}^{2}=0.071$ according to ridge regression, $Q_{GE|Metab}^{2}=0.090$ for lasso regression). The final estimated combined predictive ability of metabolites and transcriptomics  is slightly larger than 40 \% ($Q_{Metab, GE}^{2}=0.415$ according to ridge regression, $Q_{Metab, GE}^{2}=0.458$ for lasso regression). In summary, our main analysis suggests that adding gene expression to an existing model based on NMR-metabolome will potentially lead to a significant improvement in the prediction of BMI at 7 years of follow-up.

The lower part of Table 3 contains the results of our secondary analysis, in which we consider gene expression as primary source and we evaluate the addition of metabolomics. As expected, we observe a better performance of NMR-metabolome than transcriptomics to predict future BMI, for both ridge and lasso regression. In contrast to metabolites, for which the estimated value of $Q_{Metab}^{2}$ was only marginally affected by the two considered regularization methods, the type of shrinkage plays a more important role when considering gene expression ($Q^{2}_{GE}=0.092$ with ridge regression, $Q^{2}_{GE}=0.277$ with lasso  regression). Accordingly, the estimated added predictive value of metabolites with respect to a primary model based on gene expression is strongly affected by shrinkage type. Namely, metabolome explains around 19\% of the remaining variation of the outcome after accounting for the ridge-regression-based prediction calibration using gene expression data, while the equivalent added predictive ability of metabolome drops to 10\% if we model gene expression with a lasso penalty in the first step. Interestingly, when considering lasso regression, the impact of the considering transcriptomics or metabolomics as primary source is small, and in fact, the resulting predictive ability after the overall process is similar for both sequential models ($Q_{Metab, GE}^{2}=0.458$, $Q_{GE, Metab}^{2}=0.453$).

The results are supported by our first supplementary analysis consisting of artificially reducing the sample size of the available DILGOM data to check the stability of the estimated added predictive ability of the considered omic sources, measured through $Q_{Metab|GE}^{2}$ and $Q_{GE|Metab}^{2}$. In Table S1 we observe that the results based on a random subsample of the data of $n=100$ individuals are similar to those obtained with the actual sample, which indicates that the estimated added predictive values shown in Table 3 are reflecting the independent contribution of the secondary source in the prediction of BMI and not the effect of recalibrating the primary source through the second one due to scarcity of data in the first stage. That seems to be the case if we consider a random subsample of the DILGOM data of $n=50$, when the added predictive ability of the secondary source is presumably overestimated due to the underestimation of the predictive ability of the primary source in the first stage of the procedure. 

Our second supplementary analysis also supports our sequential approach. Table S2 shows the estimated values of $Q_{\mathbf{Metab,GE}}^{2}$ for two models in which simultaneously model both sources of predictors (without and with scaling). These results show that stacking both omic sources is a bad strategy, even with previous scaling of the two sources of predictors to avoid the impact of the different scale of metabolomics and transcriptomics. For both ridge and lasso specifications, we observe that the joint models based on stacking transcriptomics and metabolomics provide lower $Q^{2}_{Metab,GE}=Q^{2}_{GE,Metab}$ values than the initial $Q^2_{Metab}$.

With regard to the alternative strategies, Table 3 shows that the use of a single cross-validation approach overestimates the role of gene expression when considered alone (`$CV_{S}$, $\lambda_{opt}$' approach provides a considerably larger $Q_{GE}^{2}$ than `$CV_{D}$, $\lambda_{opt}$'), while the impact on $Q_{Metab}^{2}$ is lower (due to its fewer number of features, 139 metabolites versus more than 7000 gene expression features). Accordingly, we also observe that the impact of overfitting the residuals is larger when the first source is gene expression. Especially for lasso regression (for which we observe the absolute largest difference between `$CV_{S}$, $\lambda_{opt}$' and `$CV_{D}$, $\lambda_{opt}$' in the first step), we see how the added predictive value of metabolome is obscured by the overfitting of the first source ($Q_{Metab|GE}^{2}=0.012$ clearly not significantly different from 0).

With respect to the over-shrinkage of the penalty parameter in the inner loop of the double cross-validation (`2CV, $\lambda_{1se}$'), it leads to a more pessimistic estimation of the summary measures in each of the steps of the procedure, particularly when using ridge regression and the case where the gene expression is the second source to be considered  ($Q^{2}_{GE|Metab}=0.014$).

\section{Summary and discussion}
In this paper, we addressed the problem of evaluating the added predictive ability of a high-dimensional omic dataset for prediction of continuous outcomes in the context of multiple and possibly correlated omic datasets.
 
We proposed a sequential method which consists of considering the vector of residuals based on the primary source of predictors as outcome when fitting a prediction model based on the secondary source of omic predictors. This is equivalent to introducing the vector of individual predictions based on the primary source as an extra covariate with fixed weight when fitting a prediction model for the original outcome based on the secondary source of omic predictors. The use of a vector of predictions (which are fitted themselves) in a subsequent prediction model requires  cross-validation to account for the uncertainty of calibrating the first source of predictors in the procedure. We have proposed several summary measures, all of them based on double cross-validation predictions. Moreover, we have introduced a permutation test to formally test for added predictive ability of the secondary source.

In our approach, the first source of predictors is prioritized. Several reasons may motivate such an asymmetric approach to combination. On the one hand, available omic sources typically differ in cost and interpretability, and hence researchers may be interested in prioritizing more economic and interpretable sources. This was the case in our real data application based on the DILGOM study. NMR metabolomics measurements are more affordable than transcriptomic profiling, and also more interpretable in the context of BMI prediction. Hence, favoring NMR measurements seems logical in our setting. Similar reasons are typically used to favor classical clinical parameters when evaluating the addition of omic sources to clinical models. One could, for instance, add a novel (biomolecular) marker set,  e.g., metabolome, to a set of clinical features such as glucose, blood pressure, and serum cholestrol, which may potentially be correlated to the metabolome markers \cite{preval1, preval2, boosting}. On the other hand, our sequential approach entails a functional relation between the first (single-source) prediction rule and the second (based on both available omic sources) prediction model, which facilitates interpretation and avoids both mathematical and philosophical problems arising from combined models which may perform worse than the primary single-omic model. As we have shown in our real data application, the approach, widely used in low-dimensional settings, consisting of stacking $\mathbf{X}_{1}$ and $\mathbf{X}_{2}$ and calibrating a new model $f(\mathbf{X}_{1}$,$\mathbf{X}_{2})$, is problematic in the multi-omic setting. Such strategy does not guarantee that $Q_{\mathbf{X}_{1},\mathbf{X}_{2}}^{2}\geq Q_{\mathbf{X}_{1}}$, i.e., the combined model often performs worse than one of the single-source models.

We have focused on the case in which external validation datasets are not available and, hence, researchers have to fit and compare the predictive performance of different models using the same set of patients (internal validation). This is a common situation in practice. Epidemiological studies, such as the Finrisk (DILGOM), expand their clinical databases by including sequences of novel (omic) biomarker measure sets, which cover different biological processes and which are obtained using  different technologies. Access to equivalent data from other studies is typically hard to obtain. Moreover, due to technical and economical reasons, these new sets of markers are typically evaluated on a reduced number of individuals of the study, leading to $n<p$ situations.

Closely connected to the difference between the internal and external validation,  is the issue of choice between standard classic (lack-of-fit) residuals and the deletion-prediction residuals employed in this paper.  Use of standard residuals may suffice when adding novel predictor data to an established and known (external) risk score,  but in general greater caution should be applied.  Nevertheless,  use of standard residuals from previous model fitting when assessing added-value still seems to be the norm in most analyses. This applies not only to biostatistics \cite{boosting},  but also to related fields such as chemometrics \cite{Martens}.   
We have shown through simulations and the analysis of the DILGOM data that use of deletion-residuals is essential to avoid substantial bias in the assessment of the added value of a secondary predictor set,  when added to a primary set using the same patient data. Failure to do so renders results essentially meaningless and non-interpretable. 

Another feature of our method is that it does not only account for the predictive capacity of the first predictor set,  but it is also dependent on model choice. Hence, our added prediction assessment is not just evaluating the predictive impact of a secondary set of measurements in its own right,  but rather the joint impact of choice of predictors,  model and estimation approach.  

Our application to real data shows that better predictions can be obtained by adding transcriptomics to a model based on NMR-metabolomics, outperforming single-omic predictions. We have also illustrated the impact of model misspecification in our approach and shown that naive approaches which ignore the different nature, size and scale of the considered source of predictors fail, providing worse results than model based on NMR-metabolomics only.

The present work may be extended in several directions. The two-step sequential approach presented for continuous outcome may also be immediately extended to other outcome types, particularly to the classification context (binary outcome) and to time-to-event data, for which generalizations of linear regularized regression in high-dimensional settings are available. Summary performance measures as $Q_{\mathbf{X}_{1}}^{2}$ and $Q_{\mathbf{X}_{1},\mathbf{X}_{2}}^{2}$ can be still derived in the binary outcome context. However, given its reliance on the residuals, the extension of the conditional $Q_{\mathbf{X}_{2}|\mathbf{X}_{1}}^{2}$ to the binary and survival contexts is not straightforward. Also, the proposed testing procedure needs major modification when considering more complex responses. 
Moreover, in the implementation of the augmented assessment method, alternative prediction rules could also be used, beyond linear regularized regression considered here. Boosting algorithms might be an interesting choice worthy of further study in the high-dimensional augmented predictive framework, as an extension of \cite{boosting}, including more complex model specifications, such as non-linear and interaction terms \cite{Tutz,Kneib, Mar}. Also, alternative models (beyond the naive stacking of omic sources) could be proposed for the combination of the high-dimensional omic sources of predictors which could be used as an alternative to our sequential approach. For example, we could consider group penalization approaches, such as group lasso \citep{Yuan} or the recently proposed group ridge \citep{Wiel}. However, it is unclear if they would adequately perform in our testing context. All these topics are left as interesting lines of future research.

\section*{Acknowledgements}
Work supported by Grant Grant MIMOmics of the European Union's
Seventh Framework Programme (FP7-Health-F5-2012) number 305280.

\newpage
\beginsupplement
\section*{Supplement A. Application to DILGOM data. Supplementary analyses}
\subsection*{Supplementary analysis 1}

We check the impact of the sample size on the estimation of the predictive ability of each of the considered sources of predictors for BMI in the DILGOM data. The sample size is reduced ($n=50$, $n=100$, $n=200$) in order to check the robustness of the results based on the sequential double cross-validation procedure.

\begin{table}[h]
  \centering
  \caption{Application to DILGOM data. Sensitivity analysis: check impact of sample size. P-values based on 1000 permutations}
    \begin{tabular}{cc|c|c|c|c}   
      \hline
      
& CV type&$Q_{Metab}^2$ & $Q_{GE|Metab}^2$&$Q_{G}^2$&p-value \\
          \hline
  
$n=246$&$\alpha=0$,$CV_{D}$,$\lambda_{opt}$ & 0.305& 0.071 & 0.415 &$<0.001$\\    
$n=100$&$\alpha=0$,$CV_{D}$,$\lambda_{opt}$ &0.092 &0.121  &0.202  &0.002\\    
$n=50$&$\alpha=0$,$CV_{D}$,$\lambda_{opt}$ &0.062 &0.053  &0.106  &0.121\\                   
  \hline
$n=248$&$\alpha=1$,$CV_{D}$,$\lambda_{opt}$ &0.343 & 0.090 &0.458  &$<0.001$\\      
$n=100$&$\alpha=1$,$CV_{D}$,$\lambda_{opt}$ &0.288 & 0.337 &0.685  &0.003\\    
$n=50$&$\alpha=1$,$CV_{D}$,$\lambda_{opt}$ &0.324 &0.017  & 0.336 &0.653\\  
  \hline
  \hline
& CV type&$Q_{GE}^2$ & $Q_{Metab|GE}^2$&$Q_{G}^2$& p-value\\
          \hline
$n=248$&$\alpha=0$,$CV_{D}$,$\lambda_{opt}$ &0.092 & 0.194 &0.327  &$<0.001$\\    
$n=100$&$\alpha=0$,$CV_{D}$,$\lambda_{opt}$ &0.116 &0.095  & 0.191 &$<0.001$\\    
$n=50$&$\alpha=0$,$CV_{D}$,$\lambda_{opt}$ &0.037 &0.051  &0.099  &0.073\\   
          \hline
$n=248$&$\alpha=1$,$CV_{D}$,$\lambda_{opt}$ &0.277 & 0.102 &0.453  &$<0.001$\\        
$n=100$&$\alpha=1$,$CV_{D}$,$\lambda_{opt}$ &0.149 & 0.123 & 0.269 &0.031\\    
$n=50$&$\alpha=1$,$CV_{D}$,$\lambda_{opt}$ &0.175 &0.546 &0.774  &$<0.001$\\   
  \hline
    \end{tabular}%
  \label{tab:addlabel}%
\end{table}%


\newpage
\subsection*{Supplementary analysis 2}
Comparison of double cross-validation sequential approach to combination models based on stacking different sources of predictors. 

\begin{table}[htbp]
  \centering
  \caption{DILGOM data. $Q^2$ based on double cross-validation ridge and lasso fits for single omic source (transcriptomics and metabolomics) models and different alternatives to jointly consider both omic sources. Single source refer to models based on transcriptomics ($GE$) and metabolomics ($Metab$) only. `Stack' refers to models based on the joint model based on stacking the two sources. `Stack+std' refers to stacking of previously scaled transcriptomics and metabolomics. Sequential refers to the sequential procedure presented in our manuscript. $Metab|GE$ considers $GE$ as primary source, while $GE|Metab$ considers metabolomics as primary source.}
    \begin{tabular}{ccccccccc}
    \hline
          &       & \multicolumn{2}{c}{Single source} && \multicolumn{4}{c}{Combination} \\
          \hline
  &&GE&Metab&&Stack&Stack+std&\multicolumn{2}{c}{Sequential}\\
  &&&&&&&$Metab|GE$&$GE|Metab$\\
  \hline
     Ridge& $Q^2$  & 0.092 & 0.305 &&0.150  & 0.168 & 0.327 &0.415   \\
   Lasso& $Q^2$  & 0.277 & 0.343 &&0.250 &0.142  & 0.453 & 0.458 \\
    \hline
    \end{tabular}%
  \label{tab:addlabel}%
\end{table}%

\newpage
\section*{Supplement B. Simulation study. Alternative approaches}
Simulation results based on two modification of the two-stage procedure presented in Section 2: $CV_{S}$,$\lambda_{opt}$ relies on single cross-validation (cross-validation is used for model choice but predictions and therefore the residuals used as outcome in the second stage are directly computed on the complete sample); $CV_{D}$,$\lambda_{1se}$ relies on  over-penalization. Specifically, instead of taking $\lambda_{opt}$ as defined in the inner loop of the double cross-validation procedure presented in Subsection 2.1., we choose a larger value for $\lambda$, namely $\lambda_{opt}+1 s.e.(\lambda_{opt})$. Simulations for these two approaches are based on the same specifications detailed in Subsection 3.1., considering $p=1000$ and $q=100$.

\begin{table}[h]
\label{TableS1}
  \centering
  {\footnotesize
  \caption{Ridge ($\alpha=0$). Mean estimates (and standard deviation in brackets) of $Q_{\mathbf{X}_{1}}^2$, $Q_{\mathbf{X}_{2}|\mathbf{X}_{1}}^2$, $Q_{\mathbf{X}_{1},\mathbf{X}_{2}}^2$  and rejection proportions of the permutation test based on  $Q_{\mathbf{X}_{2}|\mathbf{X}_{1}}^{2}$ along 500 Monte Carlo trials. $p=1000$, $q=100$.}
 \begin{tabular}{cc|c|c|c|c}
    \\   
      \hline
    {\footnotesize Scenario} & CV type&$Q_{\mathbf{X}_{1}}^2$ (Step 1)    & $Q_{\mathbf{X}_{2}|\mathbf{X}_{1}}^2$ (Step 2)& $Q_{\mathbf{X}_{1},\mathbf{X}_{2}}^ 2$ (Global)&{Rej. Prop.} \\   
          \hline 
             & & &&&\\   
                &$n=50$, $CV_{D}$,$\lambda_{1se}$ &0.70 (0.13) &0.02 (0.07)&0.73 (0.09)&0.048\\  
      1a      &$n=100$, $CV_{D}$,$\lambda_{1se}$ &0.76 (0.09) &0.00 (0.03)&0.77 (0.07)&0.030\\
            &$n=50$, $CV_{S}$,$\lambda_{opt}$ &0.95 (0.03)&0.06 (0.12)&0.97 (0.00) &0.014  \\
                 &$n=100$, $CV_{S}$,$\lambda_{opt}$ &0.92 (0.02)&0.02 (0.03)&0.98 (0.01) &0.016  \\
    & & &&&\\
          &$n=50$, $CV_{D}$,$\lambda_{1se}$ &0.09 (0.08) &0.00 (0.01) &0.10 (0.07)&0.042\\
          1b   &$n=100$, $CV_{D}$,$\lambda_{1se}$ &0.16 (0.09) &0.00 (0.00) &0.16 (0.09)&0.074\\ 
            &$n=50$, $CV_{S}$,$\lambda_{opt}$ &0.87 (0.16) &0.04 (0.11)&  0.98 (0.10)&0.048\\
              &$n=100$, $CV_{S}$,$\lambda_{opt}$ &0.69 (0.13) &0.01 (0.03)&  0.94 (0.05)&0.018\\
 & & &&&\\   
     &$n=50$, $CV_{D}$,$\lambda_{1se}$ & 0.79 (0.04) &0.00 (0.02)&0.79 (0.04)  &0.034\\
           1c   &$n=100$, $CV_{D}$,$\lambda_{1se}$ & 0.84 (0.02) &0.00 (0.01)&0.84 (0.02)  &0.056\\
             &$n=50$, $CV_{S}$,$\lambda_{opt}$ & 0.97 (0.01) &0.04 (0.10)&0.99 (0.00) &0.014\\
            &$n=100$, $CV_{S}$,$\lambda_{opt}$ & 0.95 (0.02) &0.02 (0.04)&0.99 (0.00) &0.012\\
 & & &&&\\
     &$n=50$, $CV_{D}$,$\lambda_{1se}$ &0.79 (0.04) &0.00 (0.02)&0.79 (0.04)&0.056\\
           1d   &$n=100$, $CV_{D}$,$\lambda_{1se}$ &0.86 (0.02) &0.00 (0.00)&0.86 (0.02)&0.044\\
             &$n=50$, $CV_{S}$,$\lambda_{opt}$ &  0.97 (0.00) &0.04 (0.10)& 0.99 (0.00)  &0.028\\
            &$n=100$, $CV_{S}$,$\lambda_{opt}$ &  0.97 (0.01) &0.01 (0.03)& 0.99 (0.00)  &0.020\\
            & & &&&\\ 
  \hline
     & & &&&\\
         &$n=50$, $CV_{D}$,$\lambda_{1se}$ & 0.04 (0.09)  &0.01 (0.02) & 0.05 (0.09) &0.088\\
           2a   &$n=100$, $CV_{D}$,$\lambda_{1se}$ & 0.02 (0.06)  &0.01 (0.01) & 0.01 (0.07) &0.073\\
            &$n=50$,$CV_{S}$, $\lambda_{opt}$ &0.75 (0.21) &0.61 (0.16) &  0.94 (0.18)&0.746\\
            &$n=100$,$CV_{S}$, $\lambda_{opt}$ &0.74 (0.12) &0.66 (0.10) &  0.96 (0.11)&0.972\\
 & & &&&\\    
    &$n=50$, $CV_{D}$,$\lambda_{1se}$ &0.61 (0.19) &0.08 (0.13)&0.71 (0.12)&0.170\\        
           2b   &$n=100$, $CV_{D}$,$\lambda_{1se}$ &0.64 (0.19) &0.01 (0.01)&0.78 (0.24)&0.258\\
            &$n=50$, $CV_{S}$,$\lambda_{opt}$ &0.94 (0.04)&0.20 (0.15)&0.99 (0.00) &0.116\\
             &$n=100$, $CV_{S}$,$\lambda_{opt}$ &0.91 (0.03)&0.14 (0.08)&0.99 (0.01) &0.498\\
 & & &&&\\  
    &$n=50$, $CV_{D}$,$\lambda_{1se}$ &0.05 (0.06) &0.01 (0.02)&0.06 (0.06) &0.100\\  
           2d   &$n=100$, $CV_{D}$,$\lambda_{1se}$ &0.08 (0.08) &0.01 (0.01)&0.08 (0.08) &0.217\\
            &$n=50$, $CV_{S}$,$\lambda_{opt}$ &0.68 (0.21)&0.14 (0.15)& 0.96 (0.10) &0.032\\
             &$n=100$, $CV_{S}$,$\lambda_{opt}$ &0.61 (0.14)&0.12 (0.08)& 0.93 (0.06) &0.202\\
             & & &&&\\ 
\hline  
\end{tabular}%
  \label{tab:addlabel}%
  }
\end{table}%

\begin{table}[h]
\label{TableS2}
  \centering
  {\footnotesize
  \caption{Lasso ($\alpha=1$). Mean estimates (and standard deviation in brackets) of $Q_{\mathbf{X}_{1}}^2$, $Q_{\mathbf{X}_{2}|\mathbf{X}_{1}}^2$, $Q_{\mathbf{X}_{1},\mathbf{X}_{2}}^2$  and rejection proportions of the permutation test based on  $Q_{\mathbf{X}_{2}|\mathbf{X}_{1}}^{2}$ along 500 Monte Carlo trials. $p=1000$, $q=100$.}
    \begin{tabular}{cc|c|c|c|c}
    \\
    
      \hline
    {\footnotesize Scenario} & CV type&$Q_{\mathbf{X}_{1}}^2$ (Step 1)    & $Q_{\mathbf{X}_{2}|\mathbf{X}_{1}}^2$ (Step 2)& $Q_{\mathbf{X}_{1},\mathbf{X}_{2}}^ 2$ (Global)&{Rej. Prop.} \\
          \hline   
            & & &&&\\
               &$n=50$, $CV_{D}$,$\lambda_{1se}$ &0.71 (0.07)  &0.05 (0.06)&0.76 (0.06)&0.068\\
           1a   &$n=100$, $CV_{D}$,$\lambda_{1se}$ &0.79 (0.04)  &0.01 (0.01)&0.81 (0.04)&0.024\\
             &$n=50$, $CV_{S}$,$\lambda_{opt}$ & 0.93 (0.04)  &0.09 (0.14)&0.99 (0.01) &0.038\\
               &$n=100$, $CV_{S}$,$\lambda_{opt}$ & 0.92 (0.03)  &0.04 (0.05)&0.98 (0.01) &0.052\\
& & &&&\\
                 &$n=50$, $CV_{D}$,$\lambda_{1se}$ & 0.05 (0.05)  &0.01 (0.03) &0.07 (0.06) &0.034\\ 
           1b   &$n=100$, $CV_{D}$,$\lambda_{1se}$ & 0.15 (0.07)  &0.00 (0.02) &0.15 (0.07) &0.036\\
            &$n=50$, $CV_{S}$,$\lambda_{opt}$ & 0.46 (0.33)   &0.05 (0.12)&0.73 (0.32) &0.046\\
              &$n=100$, $CV_{S}$,$\lambda_{opt}$ & 0.51 (0.17)   &0.02 (0.06)&0.83 (0.11) &0.054\\
& & &&&\\
                 &$n=50$, $CV_{D}$,$\lambda_{1se}$ & 0.73 (0.06) &0.03 (0.06)&0.75 (0.07)&0.078\\   
           1c   &$n=100$, $CV_{D}$,$\lambda_{1se}$ & 0.83 (0.03) &0.01 (0.02)&0.84 (0.03)&0.074\\
            &$n=50$, $CV_{S}$,$\lambda_{opt}$ &0.96 (0.02) &0.08 (0.15)&1.00 (0.00)&0.012\\
            &$n=100$, $CV_{S}$,$\lambda_{opt}$ &0.95 (0.02) &0.03 (0.05)&1.00 (0.01)&0.042\\
& & &&&\\
          &$n=50$, $CV_{D}$,$\lambda_{1se}$ & 0.60 (0.07)  &0.03 (0.05)&0.61 (0.07)&0.070\\
                  
           1d   &$n=100$, $CV_{D}$,$\lambda_{1se}$ & 0.78 (0.04)  &0.01 (0.02)&0.78 (0.04)&0.103\\
            &$n=50$, $CV_{S}$,$\lambda_{opt}$ &0.96 (0.00)&0.10 (0.15)&0.99 (0.00)  &0.036\\
            &$n=100$, $CV_{S}$,$\lambda_{opt}$ &0.96 (0.01)&0.03 (0.06)&0.99 (0.01)  &0.024\\
  & & &&&\\
  \hline
  & & &&&\\
     &$n=50$, $CV_{D}$,$\lambda_{1se}$ &0.41(0.11) &0.17 (0.12)&0.45 (0.13)  &0.782\\
       2a   &$n=100$, $CV_{D}$,$\lambda_{1se}$ &0.50 (0.08) &0.28 (0.09)&0.86 (0.05)  &1.000\\
           &$n=50$, $CV_{S}$,$\lambda_{opt}$ & 0.71 (0.14)  & 0.50 (0.16)& 0.98 (0.03)&0.614\\
            &$n=100$, $CV_{S}$,$\lambda_{opt}$ & 0.71 (0.08)  & 0.59 (0.09)& 0.97 (0.01)&0.958\\
 & & &&&\\
            &$n=50$, $CV_{D}$,$\lambda_{1se}$ & 0.67 (0.08) &0.08 (0.07)&0.76 (0.07)  &0.130\\                  
           2b   &$n=100$, $CV_{D}$,$\lambda_{1se}$ & 0.75 (0.05) &0.05 (0.05)&0.92 (0.02)  &0.424\\
             &$n=50$, $CV_{S}$,$\lambda_{opt}$ &0.91 (0.05) &0.20 (0.19)&0.99 (0.01)&0.072\\
             &$n=100$, $CV_{S}$,$\lambda_{opt}$ &0.90 (0.04) &0.13 (0.13)&0.98 (0.01)&0.380\\
 & & &&&\\
           &$n=50$, $CV_{D}$,$\lambda_{1se}$ &0.03 (0.05)  & 0.02 (0.05)& 0.05 (0.07) &0.098\\                 
           2c   &$n=100$, $CV_{D}$,$\lambda_{1se}$ &0.10 (0.06)  & 0.01 (0.02)& 0.10 (0.06) &0.283\\
             &$n=50$, $CV_{S}$,$\lambda_{opt}$ &0.36 (0.32)  & 0.10 (0.17) &0.68 (0.32)&0.106\\
             &$n=100$, $CV_{S}$,$\lambda_{opt}$ &0.42 (0.19)  & 0.09 (0.10) &0.54 (0.12)&0.257\\
 & & &&&\\  
\hline  
 
    \end{tabular}%
  \label{tab:addlabel}%
  }
\end{table}%

\end{document}